# The dependence of halo clustering on subhalo anisotropy and planarity

Nathaniel P. Johnson [1★] and Andrew R. Zentner [1,2★]

[1]*Department of Physics and Astronomy, University of Pittsburgh, Pittsburgh, PA 15260, USA*
[2]*Pittsburgh Particle Physics, Astrophysics, and Cosmology Center (PITT PACC), University of Pittsburgh, Pittsburgh, PA 15260, USA*

Accepted 2026 March 24. Received 2026 March 20; in original form 2025 October 9

## ABSTRACT

We show that host cold dark matter (CDM) haloes cluster in a manner that depends upon the anisotropy and planarity of their subhaloes, indicating an environmental dependence to subhalo anisotropy and planarity. The spatial distributions of both satellite galaxies about central galaxies and subhaloes about host haloes have been subjects of interest for two decades. Important questions include the degree to which satellites are distributed anisotropically about their hosts or exhibit planarity as well as the degree to which this anisotropy depends on the environment of the system. We study the spatial distributions of subhaloes in a cosmological *N*-body simulation. We find that CDM subhaloes are distributed in a manner that is strongly anisotropic and planar, in agreement with prior work, though we present this in a new way. The more novel result is that anisotropy has an environmental dependence. Systems with subhaloes that exhibit *less* (*more*) anisotropy, *less* (*more*) planarity, and reside *further* from (*closer* to) their host centres cluster more *strongly* (*weakly*). Moreover, these clustering effects are not primarily the result of the correlation between subhalo anisotropy and/or planarity and another *single* halo property upon which clustering is already known to depend (e.g. concentration). This is a new and distinct effect. We discuss the impact of this result on the anisotropies of satellites as predicted by CDM, its testability, and its possible relation to the anisotropies observed about the large galaxies of the Local Group. In an appendix, we clarify our construction of ellipsoidal mock halo catalogues.

## 1 INTRODUCTION

In the standard cold dark matter (CDM) model galaxies form within the potential wells of approximately virialized haloes of dark matter (e.g. S. D. M. White & M. J. Rees 1978; G. R. Blumenthal et al. 1984). Simulations of structure formation in the CDM paradigm show that dark matter haloes are filled with smaller, self-bound subhaloes (e.g. S. Ghigna et al. 1998; A. Klypin et al. 1999b, a; B. Moore et al. 1999). These subhaloes are thought to provide the potential wells within which satellite galaxies should form. Satellite galaxies are therefore expected to reside within their own haloes, which can be identified and examined using a variety of techniques including the study of kinematics and gravitational lensing. Research on both subhaloes and their satellite galaxies has grown at a remarkable pace since these early investigations. The study of these substructures supports not only the understanding of groups, clusters, and structure formation in general, but the smallest satellite galaxies test the limits of our knowledge of galaxy formation, dark matter, and even the initial conditions for structure formation (see J. S. Bullock & M. Boylan-Kolchin 2017, for a review).

Among the many aspects of subhaloes and satellite galaxies which have been scrutinized, the degree to which their spatial and/or orbital anisotropy may or may not be consistent with one another is an open question. P. Kroupa, C. Theis & C. M. Boily (2005) catalyzed work in this area when they claimed that the disc-like distribution of the Milky Way's (MW) classical dwarf satellite galaxies could not be consistent with CDM. However, P. Kroupa et al. (2005) assumed that CDM predicted a population of subhaloes that was isotropic about their host haloes. A. R. Zentner et al. (2005b) and N. I. Libeskind et al. (2005) pointed out that CDM does *not* predict an isotropic subhalo population. Subhaloes are distributed anisotropically and exhibit planarity. More massive and earlier-forming subhaloes exhibit stronger anisotropy/planarity than the general subhalo population (A. R. Zentner et al. 2005b; A. R. Zentner 2006). Furthermore, when subhaloes thought to host large satellite galaxies are selected, the preference for planarity increases. However, the possible resolution proposed by A. R. Zentner et al. (2005b) and N. I. Libeskind et al. (2005) raised a different issue. In order to explain the orientations of the satellites in the MW and M31, the angular momenta

★ E-mail: npj16@pitt.edu(NPJ), zentner@pitt.edu(ARZ)

of these disc galaxies must likely be misaligned with the angular momenta of their haloes.[1]

The evidence that the satellites of the MW and M31 exhibit a planar distribution has expanded significantly over the past two decades and now includes the fact that the observed satellites orbit coherently with nearly coincident orbital poles (e.g. M. S. Pawlowski et al. 2012; R. A. Ibata et al. 2013; S. T. Sohn et al. 2017; M. S. Pawlowski 2018; I. M. Santos-Santos, R. Domínguez-Tenreiro & M. S. Pawlowski 2020a). Moreover, such planar structures may also characterize satellite systems other than those of the MW and M31 (R. B. Tully et al. 2015; O. Müller et al. 2018, 2021; M. S. Pawlowski et al. 2024). These additional findings include alignment of subhalo positions and orbits in phase space and pose a specific challenge. Meanwhile, a number of theoretical studies argue that while such planar and coherent configurations of satellites are not typical, they occur with sufficiently high frequency in simulations of structure formation that current observations do not yet contradict the CDM picture (e.g. T. Sawala et al. 2016; I. Santos-Santos et al. 2020b; J. Samuel et al. 2021; T. Sawala et al. 2023; P. U. Förster et al. 2022; K. Pham, A. Kravtsov & V. Manwadkar 2023; N. Garavito-Camargo et al. 2024; C. Hu & L. Tang 2025; M. Gámez-Marín et al. 2025). On the other hand, several authors argue that planes that are as thin and coherent as those observed are rare enough in cosmological simulations that current observations already contradict CDM predictions (e.g. M. S. Pawlowski & P. Kroupa 2020; M. S. Pawlowski et al. 2024; C. Seo et al. 2024; K. J. Kanehisa, M. S. Pawlowski & N. Libeskind 2025).[2] Consensus has not been reached on whether or not the degree of anisotropy/planarity of predicted and observed satellite galaxies are consistent with one another and the situation remains unresolved.

The degree to which subhaloes or satellite galaxies are anisotropic can have consequences that are significantly broader than studies of local galaxies. Anisotropy can be detected statistically in contemporary, precision galaxy clustering measurements (e.g. M. Azzaro et al. 2007; I. Agustsson & T. G. Brainerd 2010; E. van Uitert et al. 2012; A. Skielboe et al. 2012; T.-h. Shin et al. 2018; P. Wang et al. 2019; T. G. Brainerd & A. Samuels 2020) and neglecting such anisotropy can lead to systematic errors in the interpretation of these data sets (B. Hadzhiyska et al. 2023; Z. Zhai & W. J. Percival 2024; S. Ortega-Martinez et al. 2025). Strong gravitational lensing is a promising probe of dark matter substructure. However, haloes which cause lensing are biased to be observed along their longest principle axis (J. F. Hennawi et al. 2007). If subhaloes are anisotropically distributed in a manner that is correlated with the halo principle axes, as proposed by A. R. Zentner et al. (2005b) and N. I. Libeskind et al. (2005) and recently emphasized in this context by L. Mezini et al. (2025), then lensing probes of substructure will extract biased subhalo populations relative to the global average predicted by CDM (Y. D. Hezaveh et al. 2016). Finally, large statistical studies of satellite and subhalo anisotropy provide a very specific test of both the CDM model and the galaxy–halo connection. Not only is it useful and necessary to understand the anisotropy of the subhalo (or satellite galaxy) distributions, but it is necessary to understand the degree to which this anisotropy itself depends upon environment. Understanding this dependence will lead to greater understanding of the formation and evolution of both dark matter haloes and the galaxies which they host.

In this paper, we begin a statistical study of the dependence that host halo environment has on the spatial configuration of subhaloes within their hosts. This can enable large-scale, statistical studies of satellite orientations to test the predictions of the standard model of cosmological structure formation. We examine a set of host haloes in a cosmological *N*-body simulations, along with their respective subhaloes, and calculate for each host halo several quantities that reflect the degree of anisotropy and planarity of their subhalo distributions. We then study the way in which these host haloes cluster as a function of the anisotropy or planarity of their subhaloes using both standard two-point correlation functions and marked correlation functions.

Our study yields several interesting results. First, we confirm that subhaloes are quite generally distributed anisotropically about their host haloes. Subhalo positions are preferentially well-aligned with the triaxial mass distributions of their host haloes. This confirms the results of a number of the earlier studies discussed above, though we present these results in a novel manner. We also present new evidence that subhaloes do not trace the ellipsoidal shape of the overall host halo mass distribution. Subhaloes are more isotropic and less planar than the overall mass distribution of the host halo (Fig. 4).

Secondly, and more the focus of the present study, we find that host haloes cluster in a manner that depends strongly upon the spatial distributions of their subhaloes (e.g. Fig. 6). This is manifest in several ways. (1) Host haloes in which subhaloes are relatively *less aligned* with the mass distribution of the host, cluster *more strongly* than those in which subhaloes are better aligned. (2) Host halo systems in which the subhaloes are distributed in a manner that is *less planar* than average also cluster *more strongly*. (3) Systems in which subhaloes lie at *larger* halocentric radial positions cluster *more strongly* than systems in which subhaloes lie at smaller halocentric radii. In other words, host haloes which have *less* spatially-concentrated subhalo distributions, cluster *more strongly* than those with more concentrated subhalo distributions. The last of these findings corroborates several previous studies using similar metrics of the radial distributions of subhaloes (H. Miyatake et al. 2016; S. More et al. 2016; A. Dvornik et al. 2017; Y.-Y. Mao et al. 2018; X. Xu & Z. Zheng 2018). While not directly related to subhalo anisotropy, this is an important aspect of subhalo distributions that is relevant to applications related to this work. Moreover, these effects are sufficiently strong as to suggest that they could be measured observationally and used to test the standard model of structure formation at a detailed level. We pursue this in a follow-up study.

Finally, we tested whether or not the dependence of host halo clustering on the spatial distributions of subhaloes are induced by correlations between subhalo spatial distributions and other secondary host halo properties upon which host halo clustering is known to depend. Examples of such properties upon which host halo clustering is known to depend are (in addition to mass), host halo concentration, angular momentum, shape, and subhalo abundance. We do not find evidence that subhalo anisotropy/planarity-dependent clustering is induced by a correlation between subhalo anisotropy/planarity and any *single* additional property (e.g. Fig. 9). While it remains possible that a multivariate correlation between anisotropy metrics and host halo properties may explain subhalo anisotropy-dependent halo clustering, this suggests that these environmental dependencies may be novel and distinct.

[1]Yet another oddity is that the metal-poor globular clusters of the MW and M31 exhibit planarity which is oriented similarly to the satellite galaxy population, even though globular clusters are not thought to form within subhaloes (F. D. A. Hartwick 2000; A. R. Zentner et al. 2005b; M. S. Pawlowski, J. Pflamm-Altenburg & P. Kroupa 2012).

[2]Though, in some instances the debate is over whether or not a $\sim 2 - 3\sigma$ tension is sufficient to warrant a ‘challenge’ to CDM. More decisive data and methods are likely needed.

This manuscript is organized as follows. In the next section, we describe our methods, including a discussion of correlation functions and marked correlation functions, the simulation which we analyse, the selection of haloes from the simulation, and the ways in which we characterize the spatial distribution of subhaloes. In Section 3, we present our primary results in detail. These include a quantification of subhalo anisotropy and halo planarity and the degree to which host halo clustering depends upon subhalo anisotropy and planarity. We discuss our findings in relation to other known host halo clustering effects in Section 4. This is where we show that the clustering dependence of subhalo spatial distributions is distinct from other known halo clustering dependencies. In Section 4, we also place our findings in the context of other work and suggest future studies. We summarize our results and draw conclusions in Section 5.

## 2 METHODS

### 2.1 Simulations and halo catalogues

We analyse the clustering of haloes as a function of their subhalo properties in a catalogue of haloes taken from a cosmological *N*-body, gravity-only, simulation of structure formation. In this section we discuss the simulations and halo catalogues used in this paper as well as the way in which we filtered the halo catalogues to produce halo samples. We present in this paper an analysis of the $z = 0$ snapshot of the Small MultiDark-Planck (SMDPL) simulation (A. Klypin et al. 2016) because it represents a useful compromise between large volume (large samples of hosts and accurate clustering) and well-resolved subhaloes. SMDPL has a simulation box side length of 400 $h^{-1}$ Mpc with $3840^3$ particles inside. The mass resolution of each particle is $9.63 \times 10^7\ h^{-1}\,\mathrm{M}_{\odot}$ and the force resolution is 1.5 $h^{-1}$ kpc. SMDPL has adopted the Planck 2013 $\Lambda$CDM cosmology (Planck Collaboration XVI 2014) as follows: $h = 0.6777$, $\Omega_\Lambda = 0.692885$, $\Omega_\mathrm{m} = 0.307115$, $\Omega_\mathrm{b} = 0.048206$, $n_s = 0.96$, and $\sigma_8 = 0.8228$. All of our halo catalogues where produced using the ROCKSTAR halo finder (P. S. Behroozi, R. H. Wechsler & H.-Y. Wu 2012) and were accessed through the cosmosim.org web interface. We have performed similar analyses on the Bolshoi-Planck and MultiDark simulations and obtained similar results but with more noise.

#### 2.1.1 *Halo filtering*

When ROCKSTAR is used to identify haloes from particle data it is important to filter the halo catalogue as necessary. Firstly, some of the haloes ROCKSTAR identifies consist of very few particles and are considered not well resolved. For this reason, we imposed a minimum subhalo mass of $6.07 \times 10^9\ h^{-1}\,\mathrm{M}_{\odot}$, or 63 particles with the simulations particle mass of $9.67 \times 10^7\ h^{-1}\,\mathrm{M}_{\odot}$.

We also introduced a minimum mass for all host haloes. This requires that all host haloes have a mass of a certain factor greater than the minimum subhalo mass of $6.07 \times 10^9\ h^{-1}\,\mathrm{M}_{\odot}$. Since it is common for more massive host haloes to have more subhaloes (e.g. A. V. Kravtsov et al. 2004; A. R. Zentner et al. 2005a), this factor therefore determines the number of subhaloes a typical host will have in our sample. Ideally, we would like each host to have as many subhaloes as possible to reduce noise in our subhalo distribution measurements, but the more subhaloes we require, the higher our minimum host halo mass will be and the number of host haloes left in the catalogue for analysis will decrease. This will, in turn, increase the noise in our clustering measurements. We balanced this compromise by requiring hosts to have masses at least 2500 times larger than the minimum subhalo mass, which corresponds to a minium host halo mass of $1.51 \times 10^{13}\ h^{-1}\,\mathrm{M}_{\odot}$. This strikes a balance between providing a sufficiently large host sample (18 441 hosts) and a good subhalo count distribution (> 95 per cent of hosts have 10 or more subhaloes, see Fig. 1).

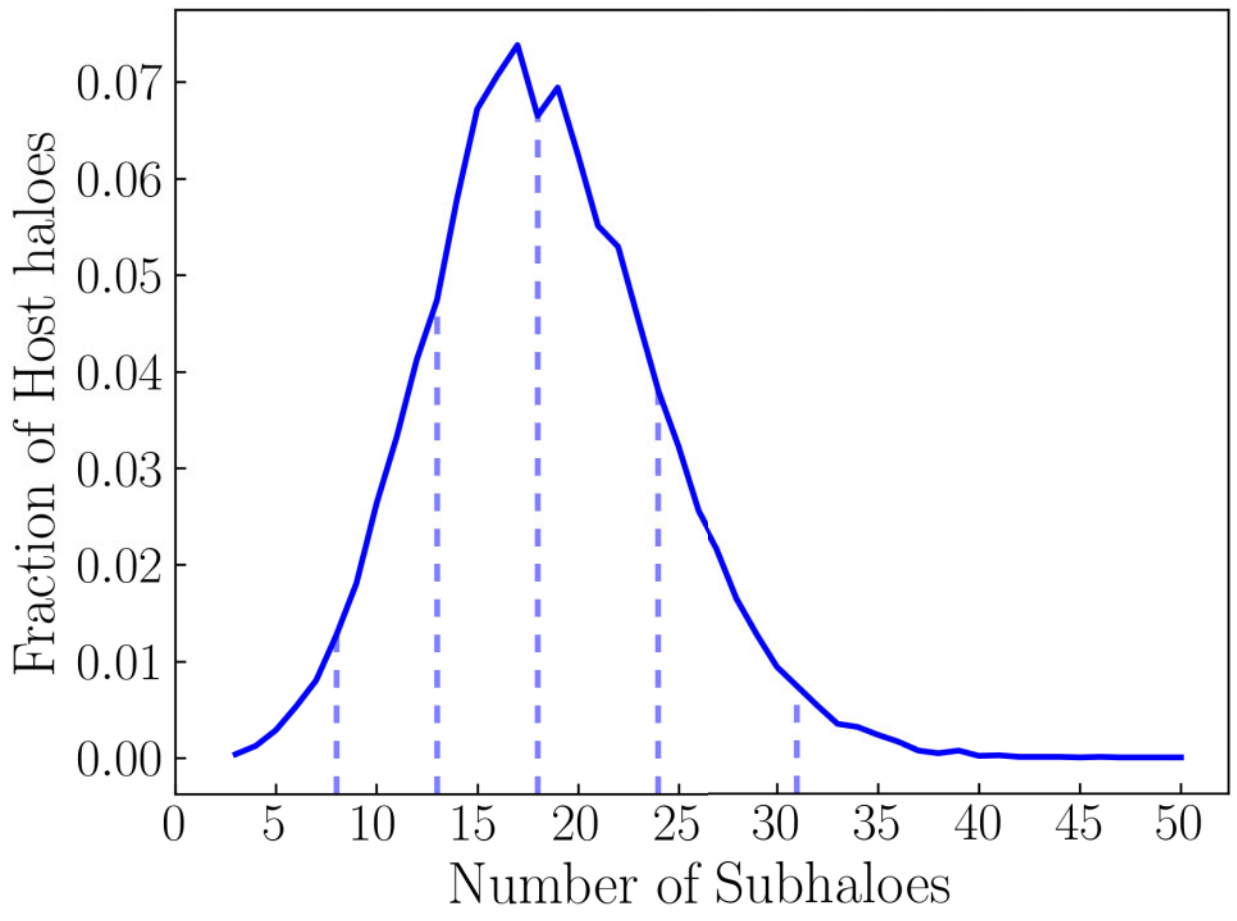

**Figure 1.** The distribution of subhalo counts following the cuts made to the catalogue as discussed in Section 2.1.1. The dashed lines show the 2.5th, 16th, 50th (i.e. median), 84th, and 97.5th percentiles of the distribution. The distribution is broad, consistent with the fact that the distribution of subhalo number at fixed halo mass is broader than Poisson (A. R. Zentner et al. 2005a; M. Boylan-Kolchin et al. 2010; C. W. Purcell & A. R. Zentner 2012; Y.-Y. Mao, M. Williamson & R. H. Wechsler 2015).

In addition to these selections, we further required every subhalo to have a mass greater than 1/2000th (0.05 per cent) of their respective host halo's mass, which reduced our subhalo population by 690 211 or 49.5 per cent. A very large number of relatively low-mass subhaloes in high-mass hosts were removed by this criterion. The purpose of this requirement is to make all host haloes have a similar number of subhaloes and exploits the fact that subhalo counts are approximately, though not exactly, self-similar across host halo masses ($N_\mathrm{sub}(M_\mathrm{sub}/M_\mathrm{host}) \propto M_\mathrm{vir}$, see A. R. Zentner et al. 2005a for a detailed discussion). This requirement removes the relation between host mass and number of subhaloes and causes all host haloes to have a similar number of subhaloes. Mass dependence of any particular halo property is further eliminated using the procedure described in Section 2.4. We have confirmed by explicit calculation that our results do not differ qualitatively when different mass cuts are made, though they do exhibit small, quantitative differences. Higher mass cuts generally result in slightly *stronger* secondary clustering biases (particularly for planarity-dependent clustering, we cannot reliably test lower mass cuts because of the resolution of the simulation). In Section 4, we further show that our measured secondary clustering dependencies are not induced solely by the clustering dependence of subhalo count per host halo.

There were also some other more trivial cuts that we made. Firstly, we only accounted for first-order subhaloes. That is to say that any halo that was a subhalo of a subhalo was discarded.

**Table 1.** Summary of the cuts made to the halo catalogue as discussed in Section 2.1.1. The top row gives the minimum mass on any subhalo that we consider. The second row gives the cut on host halo mass. The third row gives the minimum value of the ratio of subhalo mass to host halo mass. This self-similarity selection renders the number of subhaloes per host approximately equal across all host masses. The fourth row gives the number of host systems that remain after our cuts and the bottom row gives the percentage of those systems which have 10 or more subhaloes within them.

| Minimum $M_{\rm sub}$ | $6.07 \times 10^9\ h^{-1}\,{\rm M}_\odot$ |
|---|---|
| Minimum $M_{\rm host}$ | $1.52 \times 10^{13}\ h^{-1}\,{\rm M}_\odot$ |
| Minimum $M_{\rm sub}/M_{\rm host}$ | 1/2000 |
| Number of Hosts After Cuts | 18 441 |
| Percent of Hosts with 10+ Subhaloes | 95.14% |

This reduced our total subhalo population by another 362 628 or 26.02 per cent. This cut makes this work more comparable to previous literature and avoids a potential ‘over-weighting’ problem in which any property that is weighted by subhalo number (such as plane thickness, see below) would be amplified by the presence of a large number of second-order subhaloes residing at similar positions. Secondly, we required all host haloes to have more than 3 subhaloes, since our measurements of subhalo distributions that involve a best-fitting plane or an inertia tensor are not well defined when there are fewer than 3 subhaloes. The cuts discussed above should make such sparse subhalo populations rare and, indeed, this reduced our host halo sample size by only 2 systems. The resulting subhalo count distribution after all cuts can be see in Fig. 1 and a summary of the cuts can be found in Table 1.

### 2.2 Correlation statistics

#### *2.2.1 Two-point correlation function*

The two-point correlation function (TPCF) is the statistic that is most frequently used to study the clustering of galaxies and/or dark matter haloes (see P. J. E. Peebles 1980). We use the TPCF in this manuscript to study the clustering of dark matter haloes as a function of the spatial distribution of their subhaloes. We compute TPCFs, $\xi(r)$, as a function of separation $r$ in real space using

$$\xi(r) = \frac{D(r)}{R(r)} - 1, \tag{1}$$

where $D(r)$ is the number of pairs of points in the data sample of interest that are separated by a distance $r$ (or in a bin of pairs labelled by separation $r$) and $R(r)$ is the number of points that are separated by a distance $r$ in a uniformly distributed mock sample of random points. In this work, all TPCFs were calculated using the HALOTOOLS python package (A. P. Hearin et al. 2017).[3]

#### *2.2.2 Marked correlation function*

In addition to the traditional TPCFs, we also study halo clustering as a function of their spatial subhalo distribution using *marked correlation functions* (see C. Beisbart & M. Kerscher 2000, for a detailed description). Marked correlation functions (MCFs) are used to analyse the way in which objects (haloes in our case) cluster as a function of a specific characteristic, often called the ‘mark.’ MCFs have a number of advantages in the study of the spatial separation of objects based on some specific property: they allow the inclusion of all data without the specification of an ‘environment’ of interest; they yield a natural method for determining statistical significance; and they do not require boundary corrections. MCFs have been used in a variety of studies (e.g. S. Gottlöber et al. 2002; A. Faltenbacher et al. 2002; R. K. Sheth & G. Tormen 2004; R. K. Sheth 2005; G. Harker et al. 2006; R. K. Sheth et al. 2006; R. A. Skibba & R. K. Sheth 2009; M. White & N. Padmanabhan 2009; R. A. Skibba et al. 2013; Y. Zu et al. 2017; A. S. Villarreal et al. 2017; S. Satpathy et al. 2019; S. D. Riggs et al. 2021; E. Massara et al. 2023; E. Mons & C. Jose 2025), but it would be reasonable to say that they remain underutilized in the study of galaxy clustering. In this paper, we adopt the form of the MCF used by R. H. Wechsler et al. (2006),

$$\mathcal{M}_m(r) = \frac{\langle m_i m_j \rangle(r) - \langle m \rangle^2}{\mathrm{Var}(m)}. \tag{2}$$

In equation (2), $m_i$ is the value of the mark assigned to the $i^{\rm th}$ halo, $\langle m \rangle$ is the mean of the marks over all haloes, $\langle m_i m_j \rangle(r)$ is the mean of the product of marks for objects separated by a distance $r$, and $\mathrm{Var}(m)$ is the variance of all the marks. Subtracting $\langle m \rangle^2$ from the covariance of the marks at a particular separation makes $\mathcal{M} = 0$ when clustering of objects is independent of the mark. The definition then scales the MCF by the variance in the mark. We use MCFs to assess the degree to which host haloes with highly anisotropic subhalo distributions may cluster differently from the overall population of host dark matter haloes. As with the TPCF, we use the HALOTOOLS python package (A. P. Hearin et al. 2017) to compute MCFs.[4]

### 2.3 Marks quantifying subhalo anisotropy, alignment, planarity, and radial distribution

To pursue a MCF analysis of the clustering of dark matter host haloes as a function of how their subhaloes are distributed, we must first define marks that quantify the spatial distribution of subhaloes about their host's centre, particularly the distribution's anisotropy. In this section, we introduce the six different characteristics we use to quantify subhalo anisotropy, planarity, and radial distribution. We further discuss the dependence of these marks on halo mass (as well as other host halo properties) and their sensitivity to finite sampling in subsequent sections. As summary of all our marks can be found in Table 2.

#### *2.3.1 Quantiles of subhalo direction cosines*

The first quantification of the anisotropic distribution of subhaloes about their hosts will use the angle between the position vector of the subhalo relative to the centre of the host and the major axis of the mass distribution of the host halo.

[3]The estimator of equation (1) is the ‘Natural’ estimator used by the `halotools.mock_observables.tpcf` method. See the HALOTOOLS documentation at https://halotools.readthedocs.io/en/latest/api/halotools.mock_observables.tpcf.html for more details.

[4]The weighted pair counts were normalized by the unweighted pair counts for the `halotools.mock_observables.marked_tpcf` method, or the ‘number_counts’ option. See the HALOTOOLS documentation at https://halotools.readthedocs.io/en/latest/api/halotools.mock_observables.marked_tpcf.html for more details.

**Table 2.** A summary of all marks introduced in Section 2.3, including the symbol used to reference each mark, the range of possible values for each mark, each mark's units, an arrow representing whether a high value (↗) or a low value (↘) of the mark means more or less anisotropy/alignment, whether the mark measures subhalo alignment with the host halo, subhalo planarity, or subhalo radial concentration, and a one sentence description of the mark.

| Mark Symbol | Range | Units | Directionality | Measure of | Description |
|---|---|---|---|---|---|
| $C_{50}/C_{90}$ | [0,1] | None | ↗ | Alignment | The 50th and 90th percentile of the direction cosine of the subhalo position vectors and host halo's major axis. |
| $D_{\rm rms}$ | $[0,\infty)$ | Med($r_{\rm sub}$) | ↘ | Planarity | The rms of the distance from all subhaloes to a plane fitted to the subhalo positions. |
| $\lvert\cos\theta_{\rm plane}\rvert$ | [0,1] | None | ↘ | Alignment | The direction cosine of the subhalo best-fitting plane's normal vector to the host halo's major axis. |
| $(c/a)_{\rm sub}$ | [0,1] | None | ↘ | Planarity | The ratio of the smallest-to-largest principal axis lengths of the subhalo inertia tensor. |
| $\lvert\cos\theta_{\rm I}\rvert$ | [0,1] | None | ↗ | Alignment | The direction cosine of the subhalo inertia tensor's major axis to the host halo's major axis. |
| Med($r_{\rm sub}$) | [0,1] | $R_{\rm vir}$ | ↘ | Concentration | The median of the subhalo radial positions relative to their host halo. Smaller values mean more radially concentrated. |

The halo catalogues that we use contain a vector specifying the direction of the major axis of each host halo (up to an overall sign), $\boldsymbol{A}_{\bf host}$. Designating the position vector of the $i$th subhalo in the host as $\boldsymbol{x_i}$, the direction cosine between these two vectors is

$$|\cos\theta_i| = \frac{|\boldsymbol{A}_{\bf host}\cdot\boldsymbol{x_i}|}{|\boldsymbol{A}_{\bf host}||\boldsymbol{x_i}|}. \quad (3)$$

The absolute value of the $\cos(\theta_i)$ accounts for the sign ambiguity in $\boldsymbol{A}_{\bf host}$. $|\cos\theta_i| = 1$ if the subhalo lies directly along the principle axis of the mass distribution of the host halo and $|\cos\theta_i| = 0$ if it lies along a vector perpendicular to the principle axis. If all subhaloes within a host were isotropically distributed about their hosts, then the distribution of $|\cos\theta_i|$ for all subhaloes would be identical to a uniform distribution ranging from 0 to 1. This quantity measures anisotropy with particular reference to the mass distribution of the host halo, making this quantity also a measure of how well the subhalo population aligns with the host halo mass distribution.

For each host halo, there is a distribution of these direction cosines which we summarize by the 50th (i.e. median) and 90th percentiles of the $|\cos\theta_i|$ values of all subhaloes contained within the host. We use these percentiles as summary statistics to quantify alignments and abbreviate these quantiles of direction cosines as $C_{50}$ and $C_{90}$ for the 50th and 90th percentiles, respectively. The geometry is represented schematically in Fig. 2 by a 2D projection of an illustrative hypothetical system.

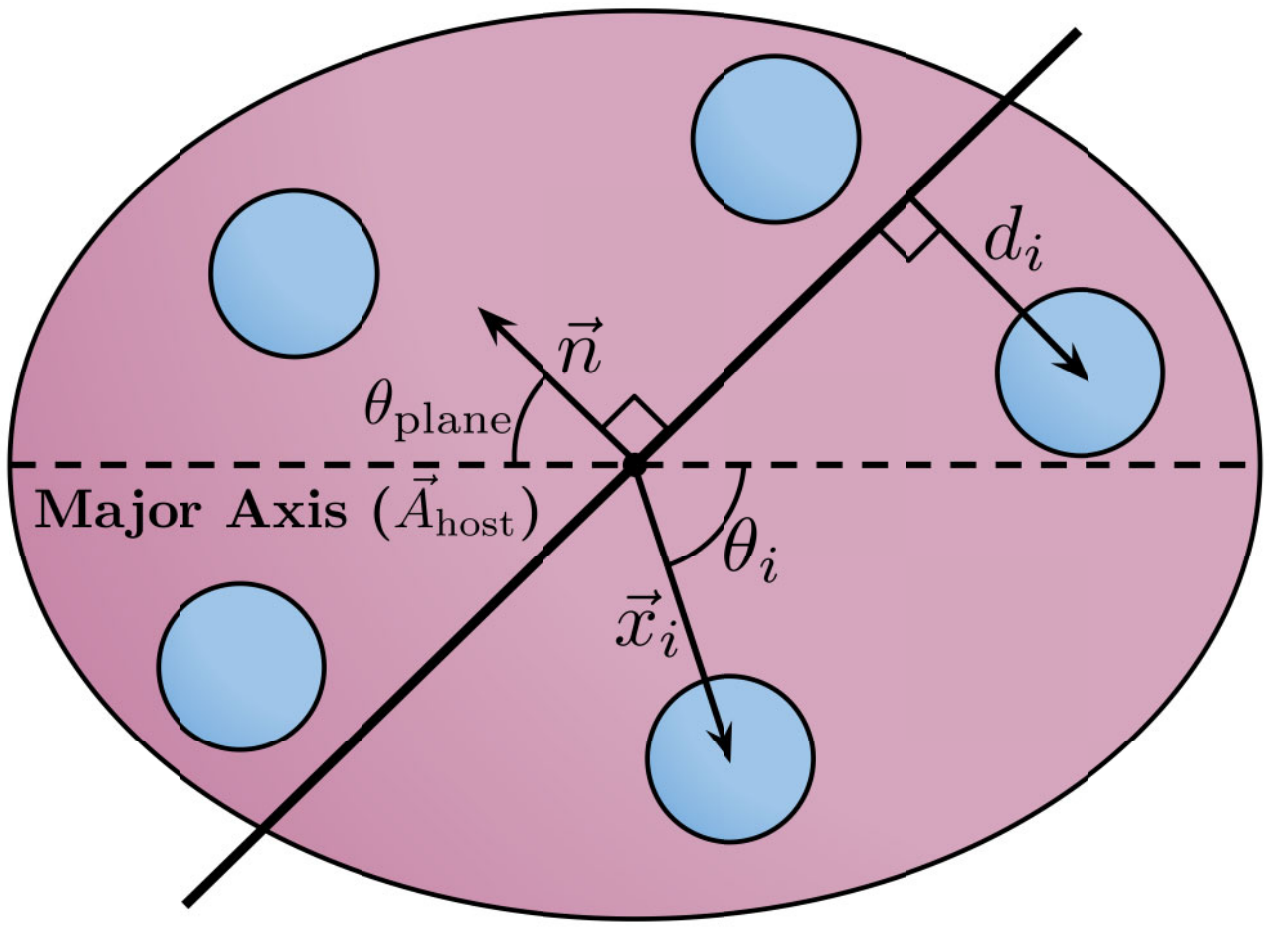

**Figure 2.** A diagram showing a 2D mock system to help explain the directional cosine, $D_{\rm rms}$, and $|\cos\theta_{\rm plane}|$ marks as introduced in Section 2.3. The large pink ellipse represents a host halo while the blue circles represent subhaloes. The dashed line visualizes the host halo's major axis, $\boldsymbol{A}_{\bf host}$, while the solid vector $\boldsymbol{x_i}$ makes the angle $\theta_i$ with the major axis (equation 3). The thick solid line spanning the diagram represents the best-fitting plane. The solid vector labelled $d_i$ is the distance from the subhalo to the plane used when determining the best-fitting plane (equation 4) and for calculating the $D_{\rm rms}$ mark (equation 5). The solid vector labelled $\boldsymbol{n}$ is the normal vector of the best-fitting plane and is used to find the angle $\theta_{\rm plane}$ when calculating the $|\cos\theta_{\rm plane}|$ mark (equation 6).

### 2.3.2 *Subhalo distribution planarity and plane thickness*

The second way in which we quantify subhalo anisotropy is with the planarity of the spatial subhalo distribution. Both subhalo and satellite galaxy planarity have been discussed using similar metrics by a number of authors (e.g. A. R. Zentner et al. 2005b; P. Kroupa et al. 2005; M. Metz, P. Kroupa & H. Jerjen 2007; M. Metz et al. 2009; P. Kroupa et al. 2010; M. S. Pawlowski et al. 2012; M. S. Pawlowski, P. Kroupa & H. Jerjen 2013; R. B. Tully et al. 2015; O. Müller et al. 2021; J. Samuel et al. 2021; K. Pham et al. 2023; L. Mezini et al. 2025)

We identify the best-fitting planes by minimizing the mass-weighted sum of the squared distances of all subhaloes to the plane,

$$\sum_{i=1}^{N} m_i d_i^2, \quad (4)$$

where $d_i$ is the perpendicular distance between the $i$th subhalo and the plane and the sum is over all $N$ subhaloes within a particular host halo. The factors of $m_i$ appearing in this equation are the subhalo masses and weight the best-fitting planes by subhalo mass. Mass weighting provides insight into the distribution of mass in subhaloes and we present results for both mass-weighted planes and number-weighted planes. To compute number-weighted planes, we simply replace all $m_i$ with 1 in equation (4) and minimize. The definition of the plane and the perpendicular distances to the plane are illustrated schematically in Fig. 2 by a 2D projection of an illustrative hypothetical system.

Once best-fitting planes are identified, we characterize their thicknesses by the root-mean-squared (rms) distance between the subhaloes within the hosts and the best-fitting planes, $D_{\rm rms}$, which is computed via

$$D_{\rm rms} = \sqrt{\frac{\sum_{i=1}^{N} d_i^2}{N}}. \quad (5)$$

Each host halo is thus assigned a single value of their subhalo plane thickness. To eliminate the dependence of the overall scale of the subhalo system in this assessment of planarity, $D_{\rm rms}$ is communicated in units of the median radial position of subhaloes in each host, Med($r_{\rm sub}$). We deemed this necessary as haloes with larger values of Med($r_{\rm sub}$) will yield larger values of $D_{\rm rms}$, and conversely for systems with smaller values of Med($r_{\rm sub}$). As has been shown by previous works (e.g. H. Miyatake et al. 2016; S. More et al. 2016; A. Dvornik et al. 2017; Y.-Y. Mao et al. 2018) and as we demonstrate further in Section 3, host haloes cluster in a manner that depends strongly upon the radial distribution of the subhaloes within them.

Larger $D_{\rm rms}$ values correspond to ‘thicker’ planes, as subhaloes typically reside further away from the best-fitting plane, which we interpret as a less planar system. On the contrary, smaller $D_{\rm rms}$ values correspond to ‘thinner’ planes, meaning the subhaloes tend to reside closer to the best-fitting plane which we interpret as a more planar system. We have computed $D_{\rm rms}$ using two distinct methods, namely with and without mass-weighting in equation (4), so we will refer to $D_{\rm rms}$ and the mass-weighted $D_{\rm rms}$ separately when discussing results.

### 2.3.3 *Subhalo plane orientation*

Our third characterization of the distribution of subhaloes about their hosts is a characterization of the orientation (rather than the thickness) of the best-fitting planes as introduced in Section 2.3.2. Defining $\boldsymbol{n}$ as the vector normal to the best-fitting plane, the cosine of the angle between the host halo principle axis and this normal vector is

$$|\cos\theta_{\rm plane}| = \frac{|\boldsymbol{A}_{\rm host}\cdot\boldsymbol{n}|}{|\boldsymbol{A}_{\rm host}||\boldsymbol{n}|}. \quad (6)$$

This measure does not reveal the degree of anisotropy of the subhalo distribution, but the degree to which it is aligned in any preferential way relative to the mass distribution of the host halo. Unlike the direction cosines discussed in Section 2.3.1, $|\cos\theta_{\rm plane}| \to 0$ when the subhalo population is well aligned with the mass distribution of the host halo. In this limit, the host halo principle axis lies within the best-fitting plane of subhaloes. Correspondingly, $|\cos\theta_{\rm plane}| = 1$ corresponds to the case in which the principle axis of the host halo mass distribution is *orthogonal* to the best-fitting plane. Fig. 2 shows this configuration qualitatively.

### 2.3.4 *Principle axis ratios of the subhalo distribution*

The fourth type of mark involves diagonalizing the inertia tensor[5] of the spatial subhalo distributions and using the principle axis ratios to characterize the anisotropy of the subhaloes. We accumulate the inertia tensor describing the subhalo distribution according to

$$\mathbf{I} = \sum_{i=1}^{N} m_i \begin{pmatrix} x_i^2 & x_i y_i & x_i z_i \\ x_i y_i & y_i^2 & y_i z_i \\ x_i z_i & y_i z_i & z_i^2 \end{pmatrix}, \quad (7)$$

where $m_i$ is the mass of the $i$th subhalo, $x_i$, $y_i$, and $z_i$ are the positions of the $i$th subhalo relative to the centre of the host halo, and the sum is over all $N$ subhaloes within the host. We diagonalize the inertia tensor to get the eigenvalues ($a^2$, $b^2$, $c^2$) which are also the squares of the principal axis lengths, where $a > b > c$ by convention. We use the axis ratio $c/a$ to characterize the anisotropy of the subhalo spatial distribution. Isotropically distributed subhalo populations would have $c/a = 1$ (as for a spherical distribution), while subhalo distributions with high anisotropy will have $c/a \ll 1$. Dark matter haloes are generally prolate, with $a > b \sim 1.1c$ (e.g. J. Dubinski & R. G. Carlberg 1991; B. Allgood et al. 2006). Moving forward, we will refer to this mark as $(c/a)_{\rm sub}$ to differentiate from the host halo’s shape, $(c/a)_{\rm host}$, which we will discuss later on in Section 4.1.

### 2.3.5 *Angle between host and subhalo distribution principle axes*

The fifth type of mark we quantify subhalo anisotropy uses the same inertia tensors from Section 2.3.4 but in a different manner. We once again start by diagonalizing $\mathbf{I}$ from equation (7) to get the eigenvectors that correspond to the principle axes of the subhalo ellipsoid, of which we are only interested in the major axis. The direction cosine between the host halo and subhalo major axes is then,

$$|\cos\theta_{\rm I}| = \frac{|\boldsymbol{A}_{\rm host}\cdot\boldsymbol{A}_{\rm sub}|}{|\boldsymbol{A}_{\rm host}||\boldsymbol{A}_{\rm sub}|}, \quad (8)$$

where $\boldsymbol{A}_{\rm sub}$ is the eigenvector corresponding to the largest eigenvalue of $\mathbf{I}$ and $\boldsymbol{A}_{\rm host}$ is still the direction of the host halo’s major axis. The absolute value in equation (8) is once again to remove any sign ambiguity from the major axis vectors. This particular mark does not quantify the degree to which the subhalo distribution is anisotropic (that is done by the principle axis ratios), but the degree to which the distribution of subhaloes is aligned with the mass distribution of the host halo. This mark was used to quantify the orientation of the subhalo population relative to the host by L. Mezini et al. (2025). For a subhalo distribution that is well aligned with the mass distribution of the host halo, we would expect $|\cos\theta_{\rm I}| \to 1$. If the subhalo distribution is not oriented in any particular way with respect to the host halo mass distribution, we should expect the values of $|\cos\theta_{\rm I}|$ to be consistent with a uniform distribution from 0 to 1.

### 2.3.6 *Median of subhalo radial positions*

The sixth and final way in which we quantitatively describe spatial subhalo configurations is a measure of subhalo radial distributions. In particular, we take the median of the subhalo radial positions, Med($r_{\rm sub}$). While this is not a measure of anisotropy, alignment, or planarity, it is interesting in its own right and it is important to mention in any study of anisotropy or planarity. As mentioned in Section 2.3.2, host haloes do cluster in a manner that depends upon the radial distribution of their subhaloes. Therefore, it is important to note and control for this effect in any measure that depends upon the radial distributions of subhaloes. For this reason, we measure plane thickness ($D_{\rm rms}$ as referred to in this work) in units of Med($r_{\rm sub}$). When discussing the Med($r_{\rm sub}$) mark on its own, however, we will measure it in units of host halo virial radius, $R_{\rm vir}$, to remove any bias in the mark due to the overall scale of the host system.

[5]This is the *shape* tensor, but is often referred to as the ‘inertia tensor’ or the ‘modified inertia tensor’ in much of the literature on halo shapes (e.g. J. Dubinski & R. G. Carlberg 1991). We follow this potentially confusing convention. This tensor defines the axis ratios of a density distribution in which isodensity contours are similar (but not confocal) ellipses.

### 2.4 The mass dependence of marks and its removal

Halo clustering has long been known to be a strong function of halo mass (N. Kaiser 1984; J. M. Bardeen et al. 1986; G. Efstathiou et al. 1988; H. J. Mo & S. D. M. White 1996). Additionally, a variety of halo properties also depend upon halo mass, including halo shape (B. Allgood et al. 2006) and, importantly, subhalo count (A. R. Zentner et al. 2005a). If the marks that we use in our study correlate with mass, then host haloes may show mark-dependent clustering due only to the underlying mass-dependent clustering. In order to study property-dependent clustering, it will then be necessary to remove the larger mass-dependent clustering signal. Fig. 3 demonstrates the relationship between mass and our marks. As is evident, our marks exhibit a small mass-dependence. Therefore, it is necessary to remove the mass-dependence of the subhalo distribution marks to ensure that the clustering dependence that we measure is not induced by the underlying mass dependence.

Following Y.-Y. Mao et al. (2018), we remove the gross mass dependence from our marks by binning the hosts by mass into bins of 100 hosts. The percentile rank of the host's true mark with respect to their bin is that host's newly assigned 'mass-normalized' version of the mark, which we will differentiate with a tilde over the assigned symbols (e.g. $\widetilde{C}_{50}$, $\widetilde{|\cos\theta_{\rm plane}|}$, and $\widetilde{(c/a)}_{\rm sub}$). We have confirmed that our mass normalization procedure, when mass itself is used as a mark, removes all mass-dependent halo clustering as desired. The color scale in Fig. 3 shows the values of the mass-normalized marks.

## 3 RESULTS

### 3.1 Subhalo anisotropy and planarity: the one-point distributions of halo marks

Before discussing the spatial relationships between host systems, we summarize the anisotropy and planarity of subhalo distributions at the population level for our host haloes. We characterize the distributions of subhaloes about their hosts using the quantities described in Section 2.3. Therefore, this is a discussion of the one-point distributions of those marks among our host halo sample.

Fig. 4 summarizes the distributions of the marks in blue. We also show the one-point distributions that the marks would have *if subhaloes were distributed isotropically* in orange. This is a useful reference because it is not clear for several of our marks the degree to which the mark distributions themselves indicate subhalo anisotropy or planarity. This can be true for a number of reasons, including the fact that our marks can be (significantly) biased by finite sampling relative to the expectations that obtain in the limit of very large samples. In addition to the isotropic mocks, we also show the distribution of marks that would be obtained if subhaloes were distributed with the same ellipsoidal shape as the underlying matter distribution. The green and red lines in Fig. 4 show the one-point distributions of the marks for these two *ellipsoidal mocks* constructed using two different methods. The ellipsoidal mocks enable us to assess the degree to which subhalo anisotropy may be determined simply by the overall ellipsoidal mass distribution.

To build the *isotropic* mock mark distributions, we built a mock catalogue based on the simulation data with isotropically distributed subhaloes and recomputed each mark for each host. In the mock catalogue, each subhalo retained its radial position, but was assigned a new angular position in a spherical coordinate system by choosing the cosine of the zenith angle, $\cos(\theta)$, from a uniform distribution on the interval $[-1, 1]$ and the azimuthal angle, $\phi$, from a uniform distribution on the interval $[0, 2\pi]$. After re-assigning the angular positions of subhaloes in this manner, all marks were recalculated. This method ensures that the simulation data and mock data are sampled the same number of times and exhibit comparable noise and bias due to finite-sampling. The resulting distributions from the isotropic mocks are shown by the orange lines in Fig. 4. The simulation data should be compared with these mock distributions in order to assess anisotropy/planarity relative to an isotropic underlying distribution. Comparison with the mock accounts for biases induced by finite sampling.

The *ellipsoidal* mocks were produced by transforming the positions of the subhaloes in the isotropic mocks in a manner that accomplishes the following objectives. (1) The subhalo inertia tensors for each system have the same principal axis ratios as the dark matter distributions within their host haloes, $[(b/a)_{\rm sub} = (b/a)_{\rm host}$ and $(c/a)_{\rm sub} = (c/a)_{\rm host}]$. We achieve this through an anisotropic stretch of the isotropic mock sample. (2) The subhalo populations within each host are rotated so that the principal axes of the subhalo distribution (the model principal axes in the limit of an infintely large sample) align with the principal axes of the host ($\hat{\boldsymbol{A}}_{\rm sub} = \hat{\boldsymbol{A}}_{\rm host}$). This is achieved through a rotation of the distribution. Of course, the *measured* principal axes of the subhalo distributions in individual systems may differ from the model due to finite sampling.

We present results from two ellipsoidal mock catalogues because there are multiple reasonable methods that can be used to build ellipsoidal mock catalogues. In our first ellipsoidal mock, which we call 'Ell. Mock $\Delta R_{\rm sub} = 0$' as a shorthand, the radial distribution of subhaloes is preserved and identical to the simulation data. In our second ellipsoidal mock, 'Ell. Mock $\rho$-iso,' the isodensity contours of subhalo density (in the limit of infinite sampling) are ellipsoids with the desired axis ratios; however, the radial distribution of subhaloes is not preserved. We explore both options because it is not apparent which is most useful and it is not possible to produce a catalogue that simultaneously matches the radial distributions of simulated subhaloes while producing ellipsoidal isodensity contours. The two methods used to construct our ellipsoidal mocks are non-trivial and their differences are discussed in detail in Appendix A. Though the two methods result in the same qualitative conclusions, we include both for completeness.

The orange lines in each panel of Fig. 4 demonstrate that the subhaloes of CDM host haloes are anisotropically distributed about their hosts. In particular, subhaloes are preferentially aligned with the major axes of mass distributions of their hosts, and distributed in a manner that is significantly more planar than would be expected from an isotropic distribution. While the panels of Fig. 4 represent these features in a somewhat novel manner, the same qualitative results have been presented in a number of previous papers on satellite distributions and anisotropy/planarity (e.g. H. Y. Wang et al. 2005; A. R. Zentner et al. 2005b; N. I. Libeskind et al. 2005; J. Bailin & M. Steinmetz 2005; A. Faltenbacher et al. 2005; A. R. Zentner 2006; N. I. Libeskind et al. 2007; N. I. Libeskind et al. 2011; M. S. Pawlowski et al. 2012; N. I. Libeskind et al. 2015; J. Shi, H. Wang & H. J. Mo 2015; X. Kang & P. Wang 2015; S. Shao et al. 2018; P. Wang et al. 2019; Y. Morinaga & T. Ishiyama 2020; K. Pham et al. 2023; J. S. M. Karp, J. U. Lange & R. H. Wechsler 2023; L. Mezini et al. 2025; K. J. Kanehisa et al. 2025).

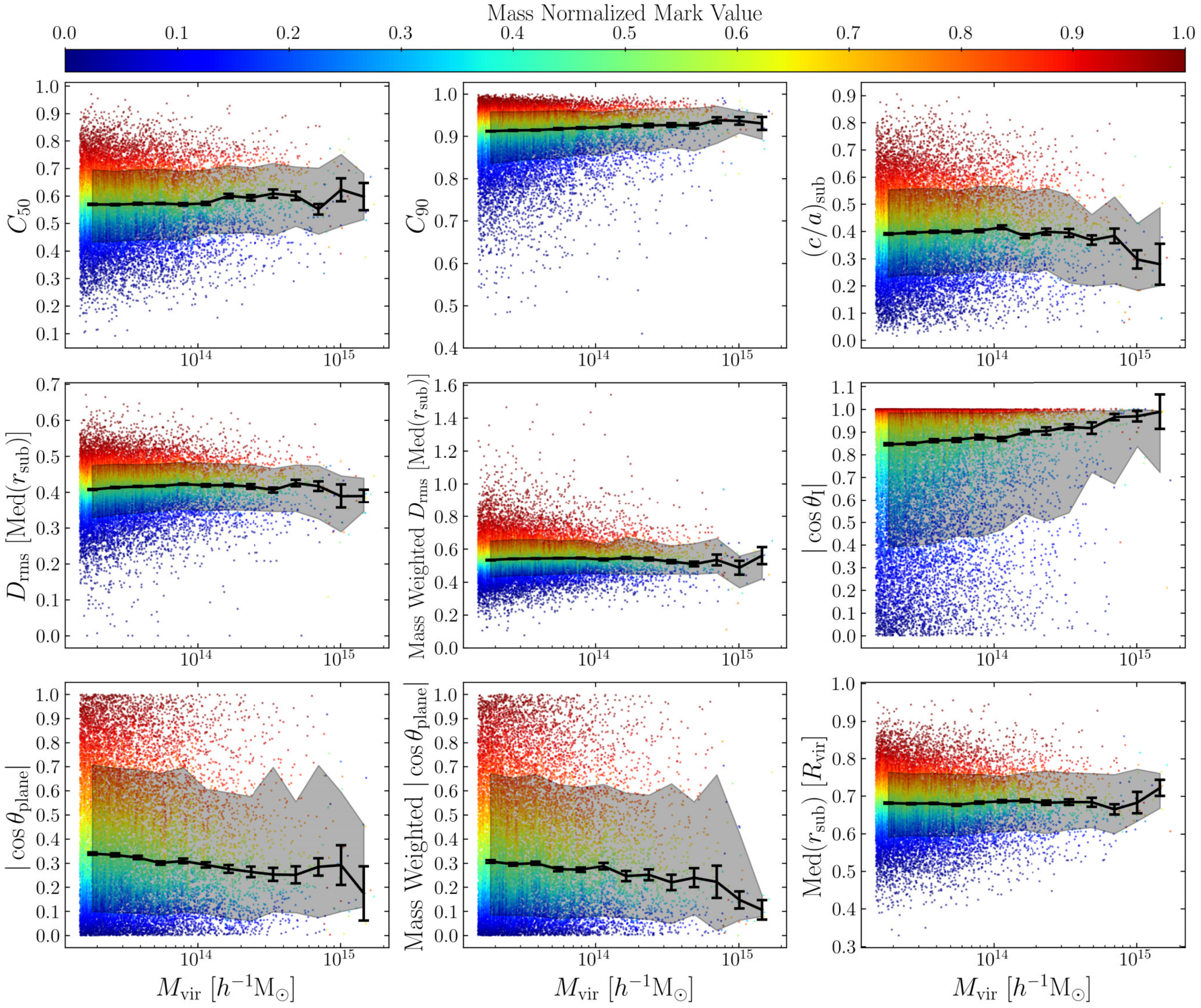

**Figure 3.** The mass dependence of subhalo spatial distribution marks and the mass normalization of our marks. Each panel shows a scatter plot of the initial mark values for each host halo (prior to mass normalization) and host halo virial mass. The black lines show the median of the initial, non-mass-normalized, marks of hosts binned by mass. The error bars show the standard error of the median for each bin. The shaded regions represents a '1$\sigma$' envelope from the 16th to the 84th percentiles of the marks in each bin. The colour of each point is determined by the new, mass-normalized value of the marks for each host. All of the new, mass-normalized marks by definition fall between 0 and 1. This figure is based off of fig. 1 of Y.-Y. Mao, A. R. Zentner & R. H. Wechsler (2018).

Next, we can examine the mock subhalo statistics computed from the ellipsoidal mock catalogues as shown in green and red in Fig. 4. The distributions of ellipsoidal mark properties are quite revealing and show that the subhaloes *do not* trace the ellipsoids of their hosts. In fact, subhaloes are distributed in a manner that is more isotropic and considerably less planar than they would be if they were tracers of the ellipsoidal shape of the host halo's underlying mass distribution. The differences in metrics are evident across all mark types and hold for both ellipsoidal mock catalogues.

Examining individual panels in Fig. 4, we see a variety of representations of the anisotropy of the subhalo population. For example, the top left and top middle panels show distributions of the $C_{50}$ and $C_{90}$ marks respectively. If subhaloes were distributed isotropically about their hosts, then $|\cos(\theta)|$ would be a uniform distribution in all cases. Therefore, unsurprisingly, the median of the $C_{50}$ mark for the isotropic distribution is $\simeq 0.5$. The median of $C_{90}$ from the isotropic mock data is slightly less than the expected value of 0.9. This discrepancy is caused by finite sampling, an effect that is included in our isotropic reference, and which is part of the reason that we use the isotropic mock to represent an isotropic hypothesis. In the limit of a very large sample, the median of the isotropic reference case does converge to 0.9, as expected. However, when sampling the subhalo distribution with only $\sim 17$ samples (see Fig. 1), measured values are biased lower. In any case, the important point to extract from the two leftmost panels in the top row of Fig. 4 is that the actual simulation data (blue) are shifted to higher values of $C_{50}$ and $C_{90}$ in comparison to the isotropic mock reference data (orange). Subhaloes are more likely to be aligned with the major axes of the mass distributions of their hosts, as recently emphasized by L. Mezini et al. (2025). As the green and red lines show, this alignment of subhalo positions with their host major axes is markedly less than it would be if the subhalo positions simply traced the triaxial ellipsoids of their hosts.

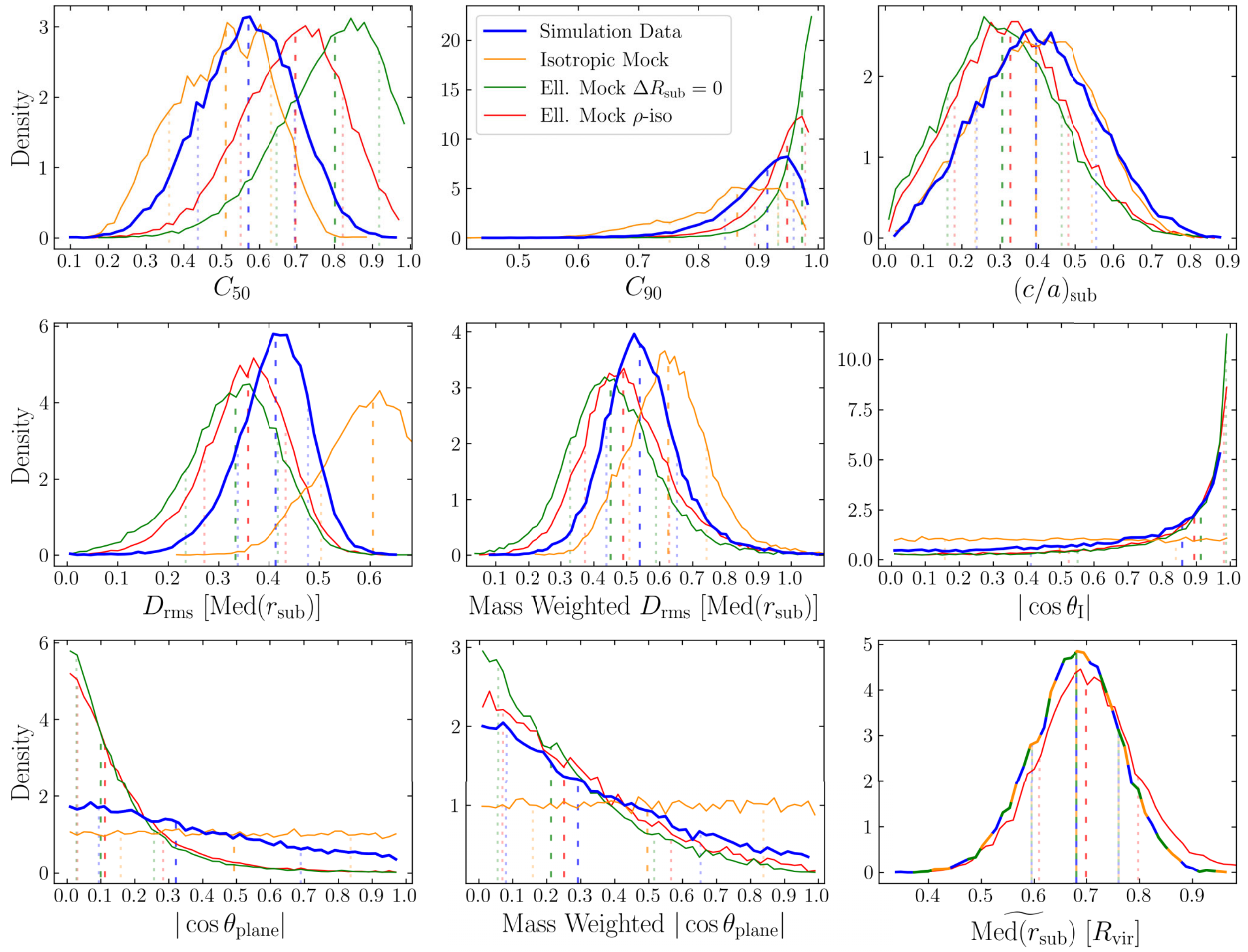

**Figure 4.** The global distributions of each mark type used in this work. These distributions are all normalized to integrate to 1. The *blue* lines show the distribution of the actual marks as calculated from the catalogue. The *orange* lines are the distributions of marks as calculated from a mock sample of the same size as the simulation data in which subhaloes were distributed isotropically about their hosts. The *green* and *red* lines are the distributions of marks when calculated using an ellipsoidal mock sample of the same size in which the subhalo distributions align with the shapes of their host haloes. The difference between the green and red distributions is the method used to calculate the ellipsoidal mock sample (neither is inherently more correct, so we present both, see Appendix A for a detailed discussion). The vertical, dashed lines shows the 50th percentile while the lighter dotted lines show the 16th and 84th percentiles of their respective distributions. Anisotropy/planarity are indicated by the differences between the simulation data and the isotropic mock data. In the case of the median subhalo radial position, the three data sets are identical. Note that we only show this comparison for our marks prior to mass normalization because after mass normalization all marks are distributed on a uniform distribution on the interval [0,1] by definition (see Section 2.4).

The top right panel in Fig. 4 shows the distribution of the $(c/a)_{\rm sub}$ mark introduced in Section 2.3.4. This figure shows only a small difference between the simulation data and the isotropic mock data. While this may seem puzzling, as $c/a = 1$ for an isotropic distribution, the reason why the median value of the mark is $\mathrm{Med}(c/a) \approx 0.4$ for the isotropic data is finite sampling. Similarly to the $C_{90}$ mark, when a large mock samples of subhaloes are used, the median of the isotropic $(c/a)_{\rm sub}$ distribution shifts toward unity as expected. This suggests that the inertia tensors defined by the subhaloes in our sample are measured with extreme noise.

The two left most panels in the second row of Fig. 4 show two measures of subhalo planarity using the $D_{\rm rms}$ mark. The left panel of the second row shows the distribution of $D_{\rm rms}$ when the planes are fit to the positions of the subhaloes and are *number* weighted and *not* mass weighted. The middle panel of the second row shows the distribution of the $D_{\rm rms}$ mark when the subhalo positions are *mass* weighted when fitting the best-fitting plane. It is apparent from these panels that subhaloes are distributed anisotropically and exhibit more planarity (thinner planes) than would be expected from a distribution that was intrinsically isotropic. The widths of the number-weighted and mass-weighted planes are $\sim 30$ per cent and $\sim 25$ per cent thinner respectively than they would be if the subhalo distributions were isotropic. In both the actual simulation data and the isotropic mock data, the mass-weighted planes are slightly thicker than the number-weighted planes, which reflects the fact that more massive haloes preferentially lie at larger halo-centric positions (e.g. D. Nagai & A. V. Kravtsov 2005; A. R. Zentner et al. 2005b; C. E. Fielder et al. 2020; L. Mezini et al. 2023). Furthermore, the plane thicknesses are significantly less in the ellipsoidal mocks (green/red) than in the actual simulation.

The rightmost panel of the middle row in Fig. 4 shows the distribution of the $|\cos\theta_{\rm I}|$ mark introduced in Section 2.3.5 for both the simulation data and the isotropic mock data. For isotropic data, $|\cos\theta_{\rm I}|$ should be uniformly distributed, as shown. The panel therefore evinces a strong alignment of subhaloes with the principle axes of the mass distributions of their host haloes in the real distribution versus the isotropic distribution (see e.g. L. Mezini et al. 2025).

The two leftmost panels in the bottom row of Fig. 4 show the distributions for the $|\cos\theta_{\rm plane}|$ mark. The left panel again shows the distribution for the number-weighted best-fitting planes and the middle panel shows the distribution for the mass-weighted planes. There is a clear shift towards increased alignment in the real distribution versus the isotropic mock distributions. Subhaloes are distributed anisotropically, and if one chooses to characterize this anisotropy by a best-fitting plane, then this plane is preferentially aligned such that the principle axis of the host halo mass distribution is close to the plane (or the normal to the plane is preferentially aligned perpendicularly to the principle axis). The difference between these two panels is clear, the mass-weighted planes are more aligned with the major axes of their host haloes, despite mass-weighted planes being thicker and less planar. This reflects the fact that the more massive subhaloes are better aligned with their host haloes than less massive subhaloes (A. R. Zentner 2006). Once again the difference between the simulation and the ellipsoidal mock is stark.

The right panel of the bottom row in Fig. 4 shows the distribution of the $\mathrm{Med}(r_{\rm sub})$ mark as introduced in Section 2.3.6. The median radial positions of subhaloes in our sample are located at $\sim 60$ per cent $-$ 80 per cent of the host halo virial radius. Despite the fact that this quantity is not a measure of anisotropy/planarity, we show results for the simulation data as well as the isotropic and ellipsoidal mock data. The distributions for the simulation data, isotropic mock, and 'Ell. Mock $\Delta R_{\rm sub} = 0$' are identical, as expected. The 'Ell. Mock $\rho$-iso' distribution shown in red is not equivalent to the other as this method for creating an ellipsoidal mock does not preserve radial position of the subhaloes, so this is expected.

This section demonstrates clearly the anisotropy of subhaloes about their hosts. Further, it is clear that this anisotropy is less than would be implied by subhaloes tracing their host ellipsoids. In the next two subsections, we present results on the clustering of host systems as a function of this anisotropy using the mass-normalized versions of these marks, as introduced in Section 2.4, to eliminate the effect of mass-dependent clustering. The mass-normalized marks are not shown here because they, by definition, will always be distributed uniformly on the interval [0,1].

### 3.2 Two-point correlation functions and visual impressions

Before proceeding to mark correlation functions, we examine the clustering of subsamples of host haloes selected by the various marks identified and studied in the preceding subsections. Three examples of this are shown in Fig. 5. In the top row, we take our sample of host haloes and split it into subsamples with $\widetilde{C}_{50}$ values above and below the median. The top left panel of Fig. 5 shows the TPCFs of the two subsamples. It is clear host haloes in which the subhaloes are *less aligned with their host principle axes* exhibit the *stronger* clustering. The top middle and top right panels of Fig. 5 make this point visually. They compare the positions of the host halo subsamples in a projection from a 125 $h^{-1}$ Mpc thick slice of the simulation. The enhanced clustering of the host haloes that have the less aligned subhalo populations is visible and is most apparent by comparing the size and frequencies of large clusters of haloes or large voids.

The middle and lower rows of Fig. 5 show similar sequences of plots. In the middle row of panels, we show that host systems with larger $D_{\rm rms}$ values cluster more strongly. That is, hosts with *less* planar systems of subhaloes cluster more strongly. The bottom row shows a parallel set of panels for $\mathrm{Med}(r_{\rm sub})$. This row of plots illustrates that host haloes in which subhaloes reside at larger halocentric radii, having less radially concentrated subhalo distributions, cluster *more* strongly. Having demonstrated that halo clustering does depend upon the spatial distributions of subhaloes, we now turn to comprehensive results using MCFs.

### 3.3 Marked correlation functions

We seek to illustrate the degree to which host haloes cluster differently dependent upon the anisotropy their subhalo populations exhibit. We summarize our results in terms of the MCFs of the marks introduced in Section 2.3. The MCFs for each of the marks we introduced are shown in Fig. 6. As with our correlation functions, all MCFs were computed using the HALOTOOLS package (A. P. Hearin et al. 2017).

The grey bands in Fig. 6 quantify the statistical significance of the measured MCFs and were constructed following C. Beisbart & M. Kerscher (2000). For each mark type, we randomly reassigned each host halo a mark from another host in the halo catalogue. This *shuffling* of the marks erases any relationship between the value of a host's mark and its position. Therefore, reshuffling creates a catalogue in which host haloes have the same one-point distribution of marks without any property-dependent clustering. We repeated this process 1000 times to create 1000 of these mock catalogues which will exhibit no property dependent clustering. The dashed grey lines give the median MCF over each of the 1000 reshuffled mock samples. The dashed lines all have $\mathcal{M}(r) \approx 0$, as expected. The filled grey band represents the band between the 2.5th percentile and the 97.5th percentile of reshuffled mocks, between which 95 per cent of the MCFs from the reshuffled mocks fall. The gray bands thus represent a '$2\sigma$' error band. In the case where there is no intrinsic property-dependent clustering, one might measure a nonzero MCF within this band for a similarly sized sample as ours. MCFs that lie outside of this gray band at multiple points (multiple values of separation distance, $r$) exhibit strong property-dependent clustering.

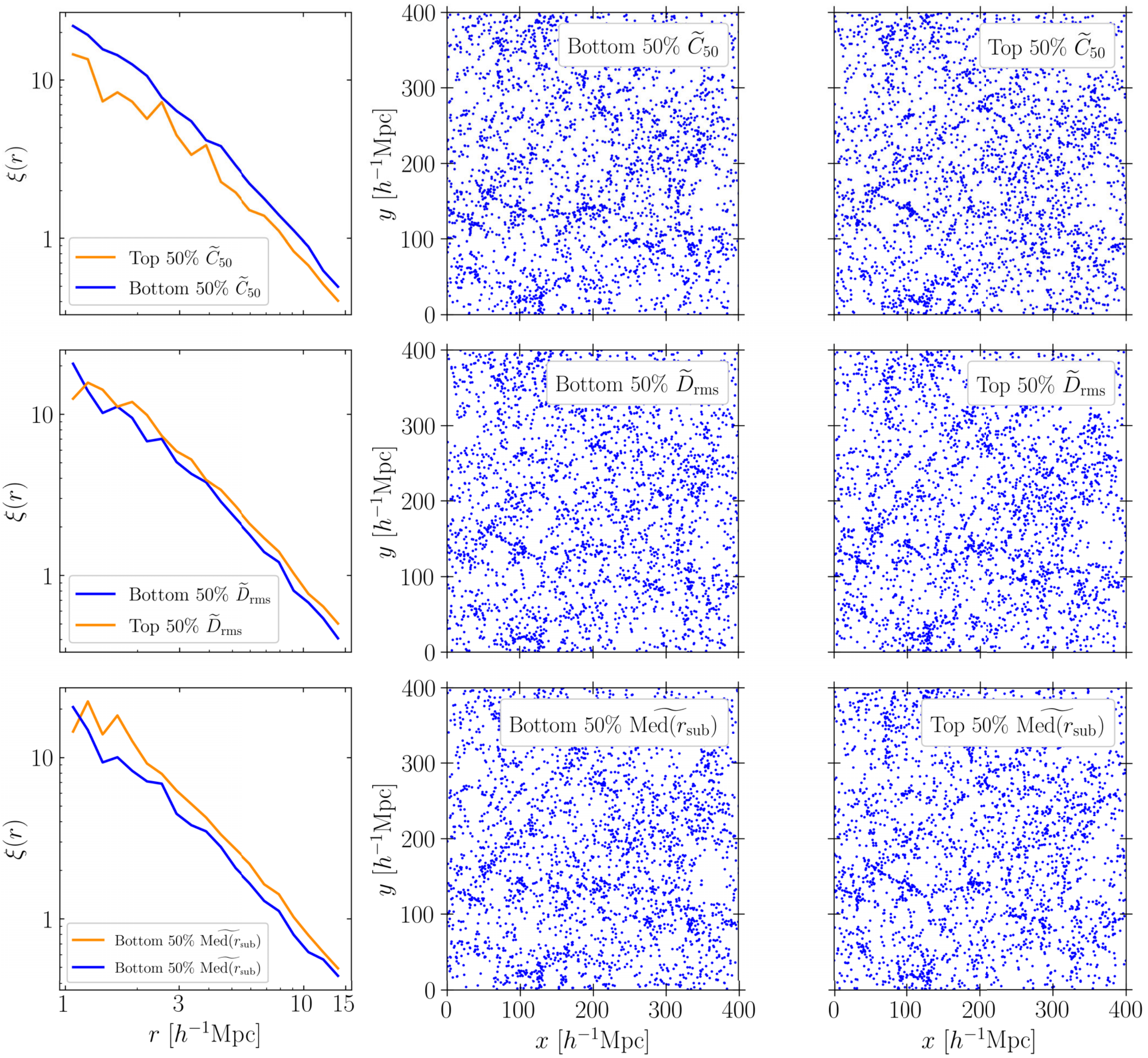

**Figure 5.** The property-dependent clustering of host haloes. This figure is organized into three rows of panels. In the top row, the left panel shows the TPCFs for host haloes in the bottom 50 per cent (blue) and top 50 per cent (orange) of subhalo–host halo alignment as measured by the mass-normalized $\widetilde{C}_{50}$ mark. The top middle and top right panels show the spatial distributions of host systems with alignment in the bottom 50 per cent (middle) and top 50 per cent (right) from a 125 $h^{-1}$ Mpc thick slice of the simulation. The other two rows of panels show a similar sequence for $D_{\rm rms}$ (middle row) and Med$(r_{\rm sub})$ (bottom row). From these panels it is clear that hosts with subhalo distributions that are less aligned with their mass distribution cluster more strongly, host with thicker subhalo planes cluster more strongly, and host systems in which subhaloes reside at larger halocentric radii cluster more strongly.

First, consider the upper left panel of Fig. 6. This panel quantify clustering as a function of how well aligned the subhalo positions are with the major axes of their host halo mass distribution, quantified by the $\widetilde{C}_{50}$ (blue) and $\widetilde{C}_{90}$ (orange) marks. There is a clear statistically significant signal in which both the sense and strength are notable. The MCFs in this panel have $\mathcal{M}(r) < 0$, implying that host haloes with subhalo distributions that are *less* aligned with the host major axis cluster *more* strongly. Moreover, not only is the effect statistically significant, but it is quite large in an absolute sense. In the case of $\widetilde{C}_{50}$, haloes in pairs separated by $\lesssim 3\,h^{-1}$ Mpc have $\widetilde{C}_{50}$ more than $\sim 0.7\sigma$ lower than the average. Moreover, haloes at all separations out to $r \sim 10\,h^{-1}$ Mpc exhibit $\widetilde{C}_{50}$ more than $\sim 0.5\sigma$ lower than the average. For $\widetilde{C}_{90}$, the magnitude of this effect is marginally smaller at small separations ($r \lesssim 5\,h^{-1}$ Mpc), but statistically significant and likewise persists at $\gtrsim 0.5\sigma$ out to $r \gtrsim 10\,h^{-1}$ Mpc.

Moving on, the middle panel in the top row of Fig. 6 show the clustering of host haloes as a function of the planarity of their subhalo populations, measured by the $\widetilde{D}_{\rm rms}$ mark for both number weighted (blue) and mass-weighted (orange) planes. As with the direction cosine marks, the property-dependent clustering signal is statistically significant. Haloes in pairs at all separations tend to have systematically *larger* thicknesses than average, whether the planes are mass weighted or number weighted. Haloes with *less* planar subhalo populations cluster more strongly than haloes with *more* planar subhalo distributions. Once again, this effect is large. Host haloes in pairs at separations from $\sim 2-10\ h^{-1}$ Mpc are $\sim 0.4-0.5\sigma$ thicker than average. This is qualitatively consistent with the results presented in the two MCFs in the top left panel of Fig. 6, namely, host haloes that contain more anisotropic subhalo populations cluster more weakly.

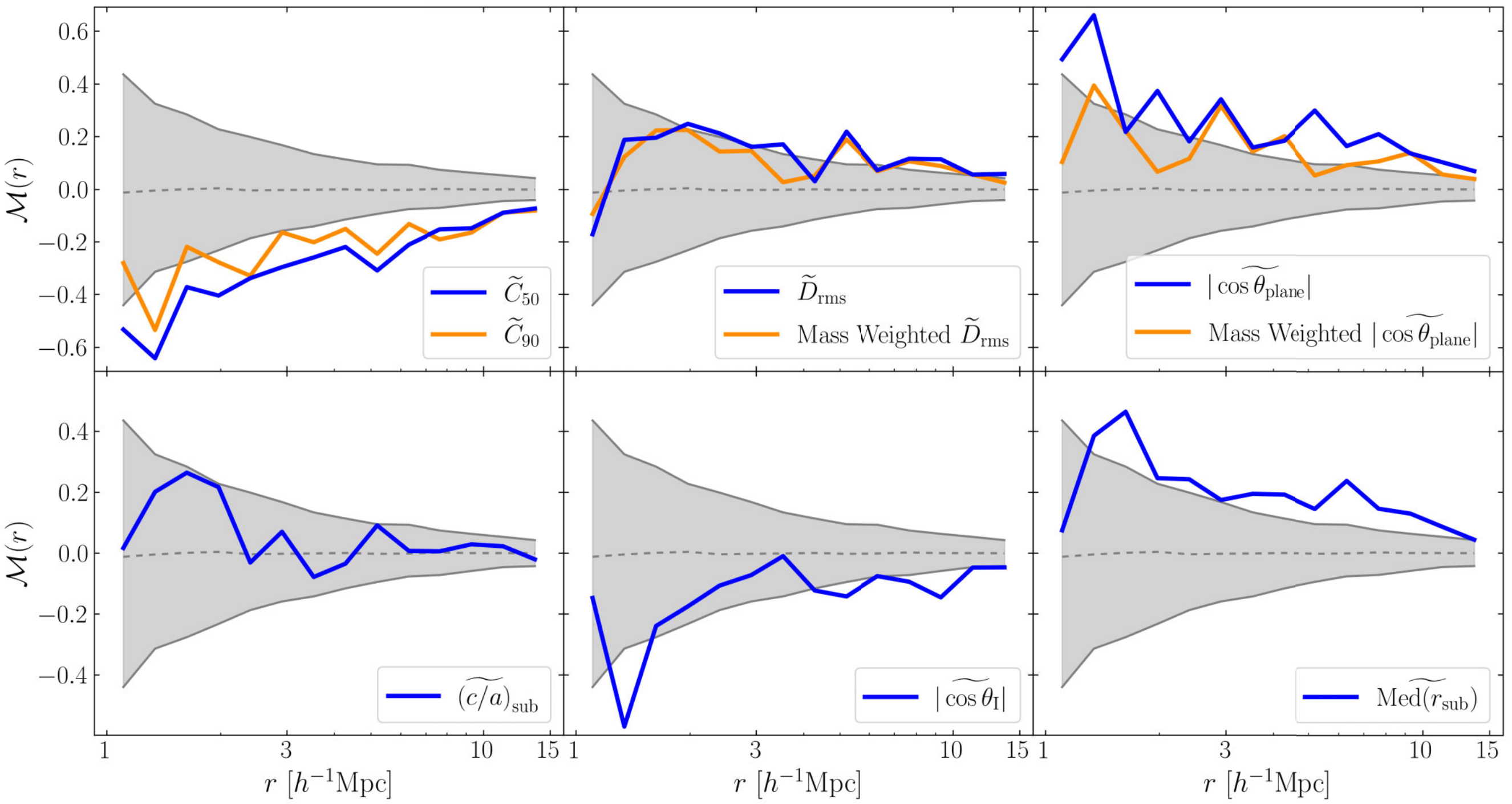

**Figure 6.** Mark correlation functions (MCFs) showing the dependence of host halo clustering on the spatial distributions of their subhaloes. The lines in each panel are the marked correlation functions using the specified mass-normalized marks indicated in the table of each panel. The shaded regions represent the middle 95th percentile of 1000 MCFs created from random permutations of the marks while the grey dashed lines are the medians of the 1000 MCFs. Thus the grey band can be thought of as a '$2\sigma$' band. MCFs that lie outside of this band at a number of points are highly statistically significant.

The top right panel of Fig. 6 show the clustering as a function of the orientation of the best-fitting plane of their satellite populations, quantified by the $|\widetilde{\cos\theta_{\rm plane}}|$ mark for both number weighted (blue) and mass-weighted (orange) planes. Recall that this mark is a measure of the cosine of the angle between the *normal* to the best-fitting satellite plane and the major axis of the host halo mass distribution. Therefore, small values ($|\cos\theta_{\rm plane}| \sim 0$) represent cases in which the plane is well aligned with the longest axis of the host halo mass distribution and larger values ($|\cos\theta_{\rm plane}| \sim 1$) represent cases in which the best fit plane of satellites is nearly perpendicular to the major axis of the host. It is clear that host systems which exhibit larger values of $|\widetilde{\cos\theta_{\rm plane}}|$ (poor alignment) cluster significantly more strongly. Again, this is qualitatively consistent with the results in the other panels of this figure.

The lower left panel in Fig. 6 displays clustering as a function of the shape of the subhalo distribution, quantified by the $\widetilde{(c/a)}_{\rm sub}$ mark. The MCF in this panel shows no statistically significant signal. However, measuring this axis ratio mark on systems in simulations is difficult because of the competing requirements of large volume (so that there are many host systems in the sample) and high resolution (which is needed to study subhaloes). Consequently, the measurements of $(c/a)_{\rm sub}$ are noisy, as discussed already in Section 3.1. The noisiness of this measurement may limit its utility as a probe of clustering in our current application.

The middle panel in the bottom row of Fig. 6 shows clustering as a function of the $|\widetilde{\cos\theta_{\rm I}}|$ mark, which is a measurement of how well the subhalo population aligns with the mass distribution of the host halo. Despite the fact that the inertia tensors of the subhaloes are measured noisily, this MCF does show a signal, although not as strong as some of our other subhalo distribution marks. It is clear that host systems in which the subhalo population aligns well with the host mass cluster weaker than average on large scales.

Finally, the bottom right of panel in Fig. 6 displays clustering as a function of the $\widetilde{{\rm Med}(r_{\rm sub})}$ mark. This mark measures the median radial position of the subhaloes in each host in units of the host virial radius. As such, it is not a measure of anisotropy, alignment, or planarity, but a summary statistic used to describe the radial distribution of subhaloes in hosts. As with the marks in most of the other panels, there is a clear property-dependent clustering signal: host haloes in which their satellites reside at larger halocentric radii cluster more strongly. This is evident at all separations and the signal is large. At a separation of $\sim 10\ h^{-1}$ Mpc, haloes in pairs have median subhalo radial positions that are $\gtrsim 0.4\sigma$ larger than typical. This aligns with previous work that investigates the clustering dependencies of similar metrics for the radial distribution of subhaloes (H. Miyatake et al. 2016; S. More et al. 2016; A. Dvornik et al. 2017; Y.-Y. Mao et al. 2018). Naively, this is contrary to the profile-dependent clustering of host haloes. It is now well known that host haloes with more concentrated density profiles cluster more strongly.

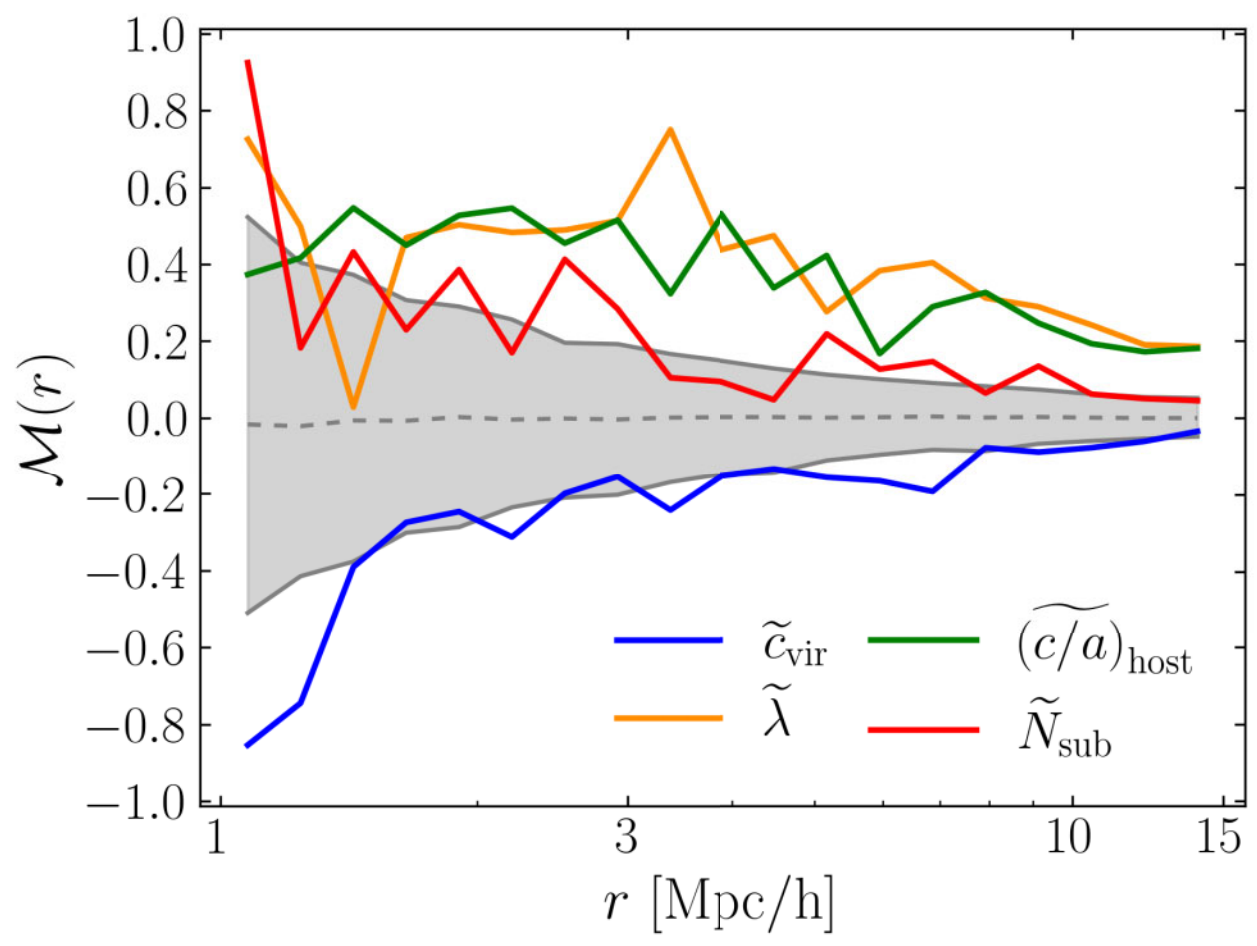

**Figure 7.** Mark correlation functions which show the dependence of host halo clustering on host halo concentration (blue), host halo spin (orange), host halo shape (green), and number of subhaloes (red). The grey lines and bands are as in Fig. 6 and represent a '$2\sigma$' statistical envelope. Each secondary bias is measured with high statistical significance across a wide range of separations. The secondary biases shown in this figure are not novel. Rather, this figure reiterates some of the known secondary biases in the literature to provide relevant context. This plot is inspired by fig. 2 in Y.-Y. Mao et al. (2018).

While clustering as a function of the radial distributions of subhaloes is not a measure of clustering which depends upon anisotropy or alignment of subhaloes, it is an observation that it is useful in the context of this study. In particular, haloes in which subhaloes are less concentrated about their hosts (distributed generally at larger halocentric radii) will also have larger best-fitting plane thicknesses, even for a fixed level of angular anisotropy. Therefore, the signal we depict in the lower right panel of Fig. 6 could give rise to clustering that depends upon subhalo plane thickness only because clustering depends upon subhalo radial position. For this reason, we measured plane thickness in units of the median subhalo position rather than subhalo position in units of the virial radius.

## 4 DISCUSSION

In the previous section, we demonstrated convincingly that host haloes in which their subhaloes are distributed more anisotropically, and in particular more well aligned with the mass of the host halo, cluster more weakly. In this section, we briefly introduce several related points of discussion including the relationship between this result and other secondary biases as well as the relevance of this work to other studies of satellite anisotropy within the Local Group and tests of $\Lambda$CDM.

### 4.1 Subhalo distribution-dependent clustering and other secondary clustering biases

Host haloes have been known to cluster as a function of a variety of halo properties for some time (e.g. L. Gao, V. Springel & S. D. M. White 2005; R. H. Wechsler et al. 2006; L. Gao & S. D. M. White 2007; A. R. Zentner 2007), as summarized conveniently in Y.-Y. Mao et al. (2018). In the previous literature, most studies focused on the clustering of haloes as a function of some measure of the formation time of the halo or a gross property of the halo mass distribution, though R. H. Wechsler et al. (2006) did point out that haloes with more subhaloes cluster more strongly. These dependencies of clustering upon halo properties other than mass are often referred to casually as 'assembly biases,' following the early papers on the subject which focused on formation time. However, they may be referred to more clearly as 'secondary biases' (mass-dependent clustering being the primary bias) (Y.-Y. Mao et al. 2018). Among the stronger and more well-studied property-dependent clustering signals, or secondary biases, are the clustering of host haloes as a function of concentration ($c_{\rm vir} = R_{\rm vir}/r_{\rm s}$), host halo angular momentum quantified by the spin parameter ($\lambda$), and host halo shape, $(c/a)_{\rm host}$.[6] The clustering of haloes as a function of formation time was the first recognized secondary bias (L. Gao et al. 2005) and now, perhaps, the most well-studied. Moreover, it is natural to suppose that anisotropy measures would be strongly correlated with measures of halo accretion history (e.g. K. Wang et al. 2020, 2025). Nevertheless, we do not show results for formation time because we find formation time-dependent clustering to be weaker than the aforementioned secondary biases for our sample of haloes.

It is therefore important to explore the connections between our subhalo spatial distribution marks (e.g. $\widetilde{C}_{50}$, $\widetilde{D}_{\rm rms}$, etc.) and these host halo properties because such connections could influence the interpretations of our results. *If* subhalo anisotropy/planarity measures are correlated with other host halo properties, such as concentration, it is possible that anisotropy-dependent or planarity-dependent halo clustering could arise solely from the known secondary biases discussed above. In that case, the clustering of hosts as a function of the spatial distribution of their subhaloes would not be a distinct effect, but would be a reflection of the correlation between subhalo spatial distribution and some other host halo property.[7]

We begin by establishing the secondary biases already described in previous literature in the sample which we select from SMDPL. Fig. 7 shows the MCFs for the 'mass-normalized' concentration ($\widetilde{c}_{\rm vir}$), spin ($\widetilde{\lambda}$), host halo shape ($\widetilde{(c/a)}_{\rm host}$), and number of subhaloes assigned to the host ($\widetilde{N}_{\rm sub}$). The subhalo numbers correspond to those assigned according to the cuts described in Section 2.1.1 and shown in Fig. 1. These halo properties have all been 'mass-normalized,' to remove the effect of mass-dependent clustering, using the methods of Section 2.4.

The *blue* line in Fig. 7 shows the MCF for host halo concentration. One might expect based on previous work high-concentration haloes to be more strongly clustered, while Fig. 7 shows that they are more weakly clustered. However, there is a significant and established mass dependence for concentration-driven clustering: high-mass haloes ($\gtrsim 10^{13}\, h^{-1}\, \mathrm{M}_\odot$, or more generally, $\gtrsim \mathrm{M}_\star$) have a negative correlation between clustering and concentration while lower-mass haloes have a positive correlation (R. H. Wechsler et al. 2006). This is depicted clearly in Y.-Y. Mao et al. (2018). Our halo sample falls clearly in the high-mass range, so it is not unexpected that Fig. 7 shows low-concentration haloes cluster more strongly. As a check of our results, we confirmed that for lower-mass haloes, high-concentration haloes cluster more strongly in the SMDPL simulation.

[6]The subscript 'host' here is used to distinguish this from the axis ratio defined by the subhalo population, $(c/a)_{\rm sub}$, discussed in previous sections.
[7]It is important to note that a correlation of a halo property A, with another property, say B, that is associated with a strong secondary halo clustering bias, does not fix the clustering of haloes as a function of property A. Haloes may or may not cluster as a function of A as expected due to the correlation of A with B. This is a consequence of the fact that correlation is not transitive. A specific example relevant to the present work is that if $\widetilde{C}_{50}$ and $\widetilde{c_{\rm vir}}$ were strongly correlated that would not be sufficient to conclude that haloes with low values of $\widetilde{C}_{50}$ cluster more strongly than haloes with larger values of $\widetilde{C}_{50}$ simply because the same is true of $\widetilde{c_{\rm vir}}$. This point has led to confusion in the literature and is discussed in detail, with examples, in Y.-Y. Mao et al. (2018).

The *orange* line in Fig. 7 exhibits the MCF of halo spin ($\widetilde{\lambda}$). Consistent with previous results, high-spin (e.g. high-angular momentum) haloes cluster markedly more strongly than low-spin haloes (e.g. P. Bett et al. 2007; I. Lacerna & N. Padilla 2012; A. S. Villarreal et al. 2017; Y.-Y. Mao et al. 2018; X. Xu & Z. Zheng 2018). In fact, haloes in pairs tend to have spins that are $\gtrsim 0.7\sigma$ larger than the typical spin of a halo of the same mass.

Continuing on, the *green* line in Fig. 7 displays the clustering of host haloes as a function of their shapes. More spherical haloes (those with *higher*$(c/a)_{\mathrm{host}}$) cluster more strongly, as has been seen in a number of earlier studies (O. Hahn et al. 2007; P. Bett et al. 2007; I. Lacerna & N. Padilla 2012; A. S. Villarreal et al. 2017; Y.-Y. Mao et al. 2018).

Lastly, the *red* line in Fig. 7 shows the dependence of host halo clustering on the number of subhaloes assigned to the host. Subhalo count is a particularly important quantity to explore in this context not only because it may be correlated with measurements of the spatial distributions of haloes, but because the subhalo count affects the fidelity with which anisotropy and planarity can be measured (this is clear in Fig. 4). As is evident, host haloes with subhalo counts that are larger tend to cluster more strongly.

We investigated the relationships between three of our subhalo distribution marks ($C_{50}$, $D_{\mathrm{rms}}$, $\mathrm{Med}(r_{\mathrm{sub}})$) and the four secondary biases we have chosen to highlight ($c_{\mathrm{vir}}$, $\lambda$, $(c/a)_{\mathrm{host}}$, $N_{\mathrm{sub}}$). We found in each combination that there is some correlation between the subhalo distribution marks and the host property marks, though the correlations may be quite weak in some instances. These relationships are shown in Appendix B for completeness, but the important point for the present discussion is that these correlations exist. With these relationships established, we now turn towards an investigation into the possibility that our measured anisotropy-dependent clustering is not a wholly unique signal but a manifestation of previously known secondary biases induced by correlations between anisotropy marks and host halo properties.

Our first test of this possibility can be seen in Fig. 8, which involves splitting our host halo sample into thirds, or tertiles, by one of the secondary halo properties and examining the clustering of each subsample. In Fig. 8, each row corresponds to one of our subhalo distribution marks, while each column corresponds to one of the secondary biases. This makes each panel a unique pairing of one of our marks and one of the secondary biases, as made clear by the text in each panel. Within each panel there are four marked correlation functions. The *black* lines show the standard MCFs for the entire host halo sample, which we have already shown in Fig. 6 and discussed in Section 2.2.2. Then the *blue* lines show the MCFs for the host haloes that have the 33 per cent lowest values of the corresponding secondary bias. The *orange* lines show the MCFs for the middle 33 per cent and the *green* lines show the MCFs for the host haloes with 33 per cent largest values. The grey shaded regions in each panel are the same $2\sigma$ error bar as in Fig. 6, calculated using MCFs from 1000 random permutations of the respective subhalo distribution marks. The lighter shaded region is the error band for the tertiles, while the darker shaded region is the error band for the standard MCF. So when comparing these MCFs, you have to compare them to their respective error bands–darker for the black lines and lighter for the colored lines. We are most keenly interested in assessing the clustering strength at larger scales, $\sim 10\, h^{-1}$ Mpc.

In the top left panel of Fig. 8, you can see the MCFs for our $\widetilde{C}_{50}$ mark when split into tertiles by host halo concentration. Since each of these MCFs show a similarly significant signal, we conclude that the $\widetilde{C}_{50}$ dependent clustering has no relation to the host halo's concentration value and therefore is independent of concentration-dependent clustering. The same can be concluded from all the panels in the top row for the rest of the $\widetilde{C}_{50}$ pairings, though the measurements on the tertiles have notably smaller signal compared to their error bands. The middle row, which corresponds to our $\widetilde{D}_{\mathrm{rms}}$ mark, exhibits more differences between the MCFs and overall more noise. This signifies that the planarity-dependent clustering we measured may be related to one or more of the known secondary biases. However, given the noisiness of the tertile measurements, this test is largely inconclusive. Finally, the bottom row shows the same set of four tests performed for $\widetilde{\mathrm{Med}(r_{\mathrm{sub}})}$. The conclusion is again clear that each tertile exhibits $\widetilde{\mathrm{Med}(r_{\mathrm{sub}})}$-dependent clustering at a level similar to the full sample to within our ability to measure. This test therefore yields no conclusive evidence that our subhalo anisotropy/planarity-dependent clustering is induced by a correlation between subhalo anisotropy and any of the individual host halo properties examined here.

A second test of the role of secondary biases on the interpretation of our results is displayed in Fig. 9. The figure contains three panels which show the MCFs of host haloes marked by $\widetilde{C}_{50}$ (left panel), $\widetilde{D}_{\mathrm{rms}}$ (middle panel), and $\widetilde{\mathrm{Med}(r_{\mathrm{sub}})}$ (right panel) computed in several ways. The *blue* lines ('Standard MCF') are the MCFs of these quantities already given in Fig. 6 and repeated here for reference. In each panel, the remaining lines show the MCFs for subhalo spatial distribution marks after removing *both* mass-dependent halo clustering *and* the clustering of haloes as a function of one additional halo property, such as concentration. The additional property is indicated by the lines above the plot. In this way, we show MCFs after removing the impact of mass-dependent clustering *and* a *single* secondary halo bias.

We removed the secondary bias by straightforward extension of the 'mass normalization' procedure described in Section 2.4 to both mass and one additional host halo property. We first binned the host haloes into mass bins each containing 500 hosts and then subdivided each of those mass bins into bins of 10 hosts according to the secondary halo property ($c_{\mathrm{vir}}$, for example). The hosts were thus sorted into two-dimensional bins and each host halo was assigned a new 'normalized mark' which is the percentile score of the subhalo spatial distribution mark inside the bin. For example, the 500 least massive host haloes are placed in the first mass bin. Then the 10 hosts in that mass bin with the smallest concentrations are placed in a concentration bin. Within that concentration bin, the percentile scores are calculated for the mark. We continue for all bins and all host haloes. In this way, both the mass-dependent clustering *and* the concentration-dependent clustering of the host haloes are removed. We performed this procedure for concentration, spin, shape, and subhalo count. We confirmed by explicit calculation that the marks so assigned yield samples which exhibit no residual mass-dependent clustering or secondary property-dependent clustering. With these new doubly-normalized marks, we computed new MCFs for our subhalo distribution marks, which are shown by the remaining lines in Fig. 9. The advantage of this approach compared to the previous test is that it does not suffer from the same loss of signal to noise, because the sample is not further subdivided.

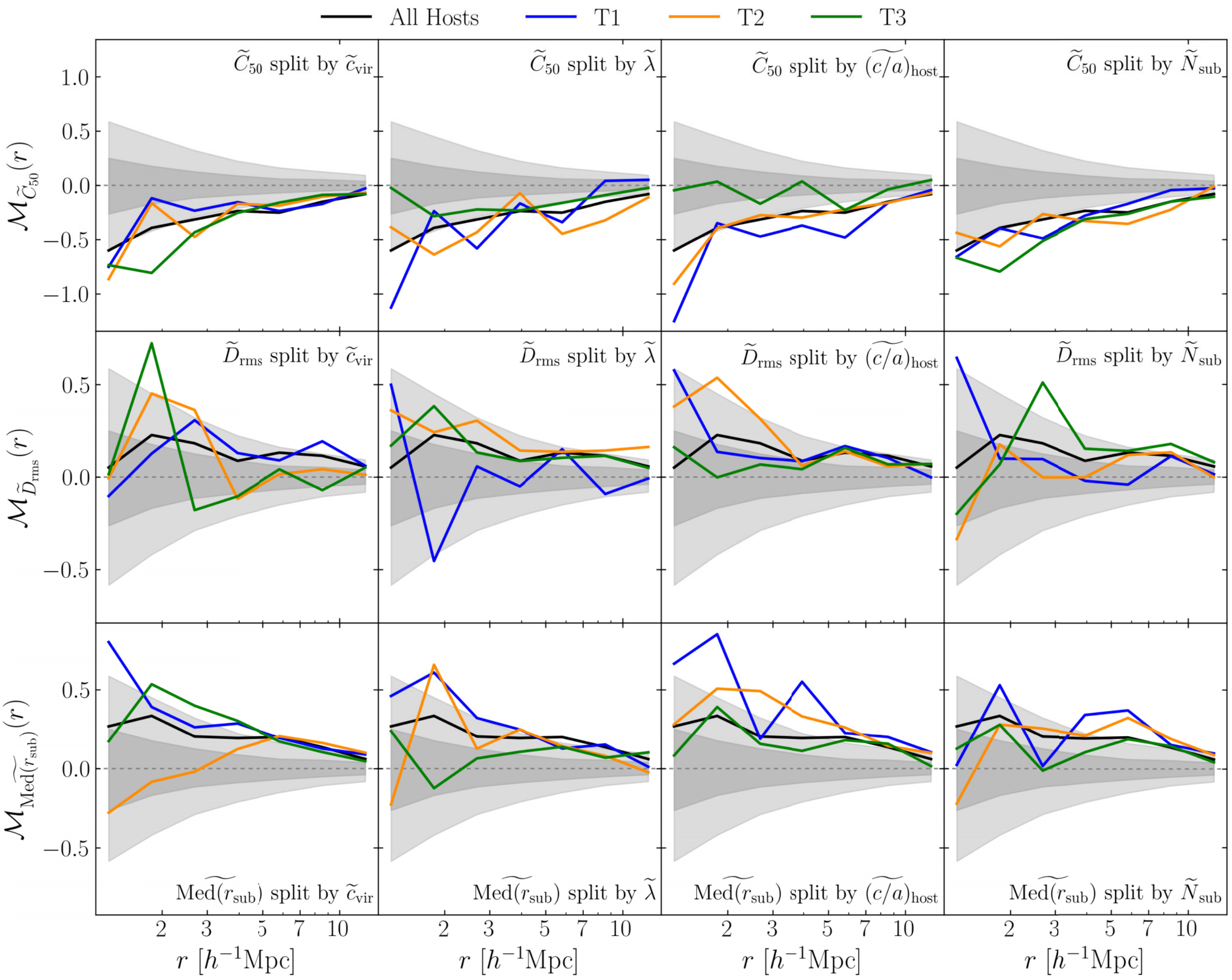

**Figure 8.** Sets of MCF demonstrating the independence of our measured subhalo anisotropy dependent clustering from the known secondary biases of $c_{\rm vir}$, $\lambda$, $(c/a)_{\rm host}$, and $N_{\rm sub}$. Each panel corresponds to a pairing of one of our subhalo distribution marks and a secondary bias mark. Within each panel there are four MCFs: *black* shows the standard MCF as already shown in Fig. 6, *blue* shows the MCF for the subsample of hosts that fall within the bottom 33 per cent of the secondary bias halo property, *orange* shows the MCF for the subsample that fall into the middle 33 per cent, and *green* shows the MCF for the subsample of haloes that fall into the top 33 per cent of the secondary halo property. The grey shaded regions are the same $2\sigma$ error bar as in Fig. 6, the lighter one being for the smaller sample size of the tertiles and the darker one being for the standard MCF. Across the top and bottom rows, all of the MCFs show the same statistically significant signal, signifying the independence of subhalo alignment and radial position dependent clustering from these secondary biases. The MCFs across the middle row show a similar qualitative signal, but to a much weaker degree, signifying that subhalo planarity is more related to these secondary biases that subhalo alignment and radial position.

The most relevant comparisons to make in Fig. 9 are to compare the *blue* lines labelled 'Standard MCF,' with the remaining lines in the same panel. The *orange* lines show a comparison to the case in which concentration-dependent secondary bias is removed, the *green* lines have spin-dependent clustering removed, the *red* lines have host shape-dependent clustering removed, and the *purple* lines have subhalo count-dependent clustering removed. Consider the leftmost panel, showing MCFs for $\widetilde{C}_{50}$. Notice that the MCFs we compute after removing secondary biases are all similar to the original MCF reported earlier in this manuscript (blue line). In fact, it is somewhat difficult to distinguish the individual lines in the left panel of Fig. 9, but this is partly the point of the exercise. This indicates that the clustering of host dark matter haloes as a function of alignment ($\widetilde{C}_{50}$) is *not* due to the correlation of alignment with halo properties such as concentration, spin, shape, or subhalo count. Alignment-dependent host halo clustering is a distinct effect. The reader can make a similar comparison for the middle ($D_{\rm rms}$) and right (${\rm Med}(r_{\rm sub})$) panels of Fig. 9, signifying that subhalo planarity and subhalo radial position dependent clustering is also a distinct effect from the secondary biases we have chosen to investigate.

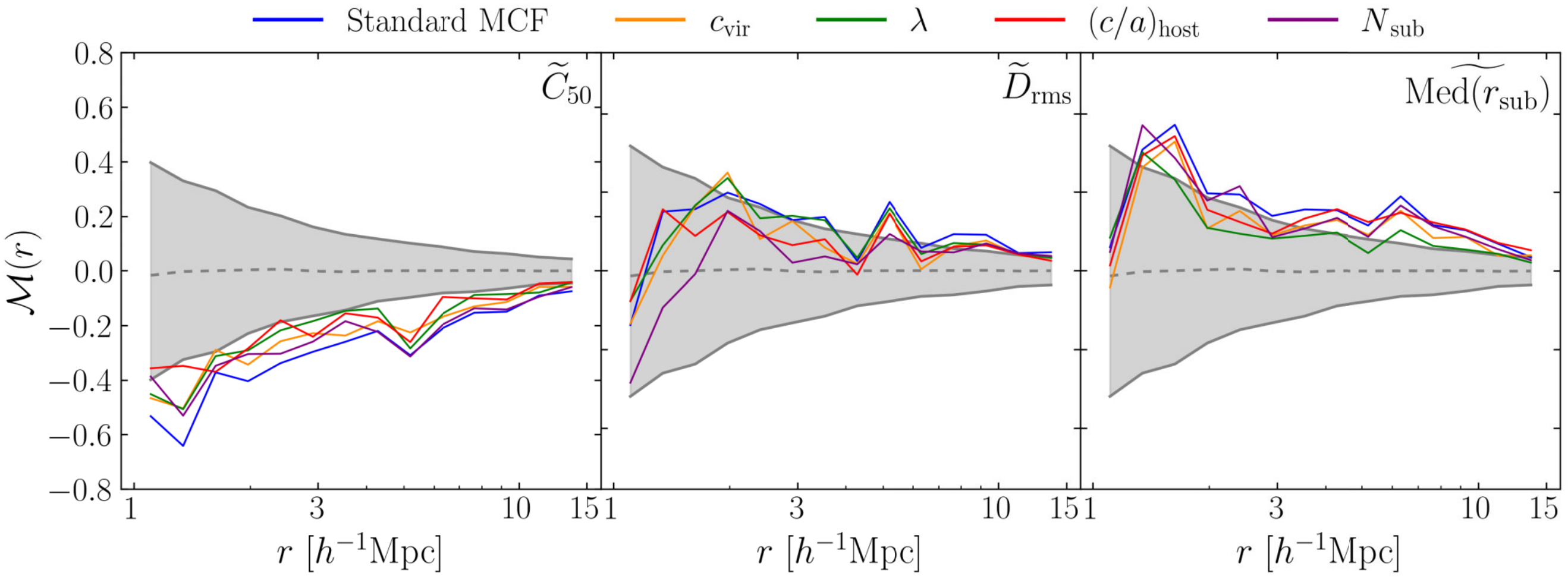

**Figure 9.** MCFs for the spatial distribution of subhaloes after removing both mass-dependent clustering and the clustering dependence of a secondary property. The blue lines simply repeat the standard MCFs from Fig. 6, for reference. The remaining lines show MCF for each spatial distribution mark that has also had a particular secondary bias removed, indicated by the colors of the lines: the *orange* lines have had concentration-dependent clustering removed, the *green* lines have had spin-dependent clustering removed, the *red* lines have had host halo shape-dependent clustering removed, and the *purple* lines have had subhalo count-dependent clustering removed. The left panel shows results for subhalo alignment, characterized by $\widetilde{C}_{50}$, the middle panel shows plane thickness ($\widetilde{D}_{\rm rms}$), and the right panel shows radial distribution ($\widetilde{{\rm Med}(r_{\rm sub})}$). Notice that in all panels, the strength of these MCF clustering signals remain statistically significant even after removal of secondary bias. This indicates that the clustering of host haloes as a function of the subhalo alignment, planarity, and radial position is not induced by the correlation of those properties with another *single* host halo property which is already known to drive halo clustering.

The MCFs with secondary bias removed shown in Fig. 9 reveals no evidence that the clustering of haloes as a function of the way in which their subhaloes are distributed is driven by a correlation between subhalo anisotropy/planarity and another *single*, known property which drives halo clustering. *The clustering of host haloes as a function of the spatial distributions of their subhaloes, particularly anisotropy and radial position, may therefore be a distinct effect not easily understandable in terms of known biases*. Of course, haloes cluster as a function of many properties. It is possible that a multidimensional analysis of a number of simultaneous correlations (such as those proposed by X. Xu & Z. Zheng 2018) could explain the subhalo spatial distribution-dependent clustering of host haloes, but this is difficult to determine with a halo sample of the size we study. We therefore leave such an exploration to future work. An additional test of these relationships can be found in Appendix C which corroborate these conclusions.

### 4.2 The local group, the holmberg effect, and ΛCDM

The last two decades have witnessed enormous interest in the anisotropy and planarity of the distribution of satellite galaxies particularly within the Local Group (e.g. D. Lynden-Bell 1982; H. Y. Wang et al. 2005; P. Kroupa et al. 2005; A. R. Zentner et al. 2005b; N. I. Libeskind et al. 2005; N. I. Libeskind et al. 2007; N. I. Libeskind et al. 2011; M. S. Pawlowski et al. 2012; N. I. Libeskind et al. 2015; J. Shi et al. 2015; X. Kang & P. Wang 2015; S. Shao et al. 2018; M. S. Pawlowski & P. Kroupa 2020; O. Müller et al. 2021; M. S. Pawlowski 2021; J. Samuel et al. 2021; J. S. M. Karp et al. 2023; L. Mezini et al. 2025; K. J. Kanehisa et al. 2025). Within the local group of galaxies, the satellite galaxies of both the Milky Way and Andromeda are observed to be anisotropic, planar, and aligned with the minor axes of their host galaxies (though the orientation of the central galaxies with respect to the principle axes of their haloes remains unknown). This effect within the Local Group is broadly referred to as the Holmberg Effect, after being described by E. Holmberg (1969). The arrangement of satellites in planes nearly coincident with the poles of their host galaxies has led to the discussion of the so-called vast polar structures of satellites (VPOS) in the Milky Way, Andromeda, and perhaps other systems (M. S. Pawlowski et al. 2012; M. S. Pawlowski 2021; R. B. Tully et al. 2015; O. Müller et al. 2021). Indeed, part of the motivation to study subhalo anisotropy statistically, as we have done, is to test with statistically-large samples the degree to which anisotropy is expected and depends upon environment. It seems sensible then to return to this motivation in closing.

For the sake of completeness and to make more explicit connection with this literature, we investigated somewhat more directly the Holmberg Effect of VPOS in the SMDPL simulation. We first defined and identified a set of systems that are analogues to the Milky Way-Andromeda (MWA) system. We defined a MWA system to be two haloes which have (1) a relative separation of between 1.5 and 2.5 times the virial radius of either host halo and (2) virial masses between 0.75 and 1.25 times the virial mass of their companion. Of course, this definition is broad, but this breadth is necessary in order to draw a sample of any significant size. This returned 332 MWA analogue systems.

Fig. 10 compares the distributions of the $C_{50}$, $D_{\rm rms}$, and ${\rm Med}(r_{\rm sub})$ marks of the entire host sample (blue) and the MWA subsample (orange). It is clear that the alignment, planarity, and radial distributions of subhaloes in all of the MWA systems we identified are similar to that of the global population, at least to within our ability to determine in a sample of this size. MWA systems according to the relatively coarse definition used here do not have subhalo distributions that are particularly out of the ordinary. This agrees with other recent results such as the similar work by K. J. Kanehisa et al. (2025), which measures clustering effects within simulations based on halo properties of subhalo lopsidedness. A more refined analysis with a more selective definition of a MWA system is not possible with this simulation set due to a combination of limited statistics and limited resolution. As we discussed in the introduction, analyses using zoom-in simulations show that significant satellite anisotropy and planarity are possible such that the VPOS and similar local observations alone do not contradict the standard ΛCDM paradigm (e.g. T. Sawala et al. 2016; I. Santos-Santos et al. 2020b; J. Samuel et al. 2021; P. U. Förster et al. 2022; K. Pham et al. 2023; T. Sawala et al. 2023; N. Garavito-Camargo et al. 2024; M. Gámez-Marín et al. 2025; C. Hu & L. Tang 2025).

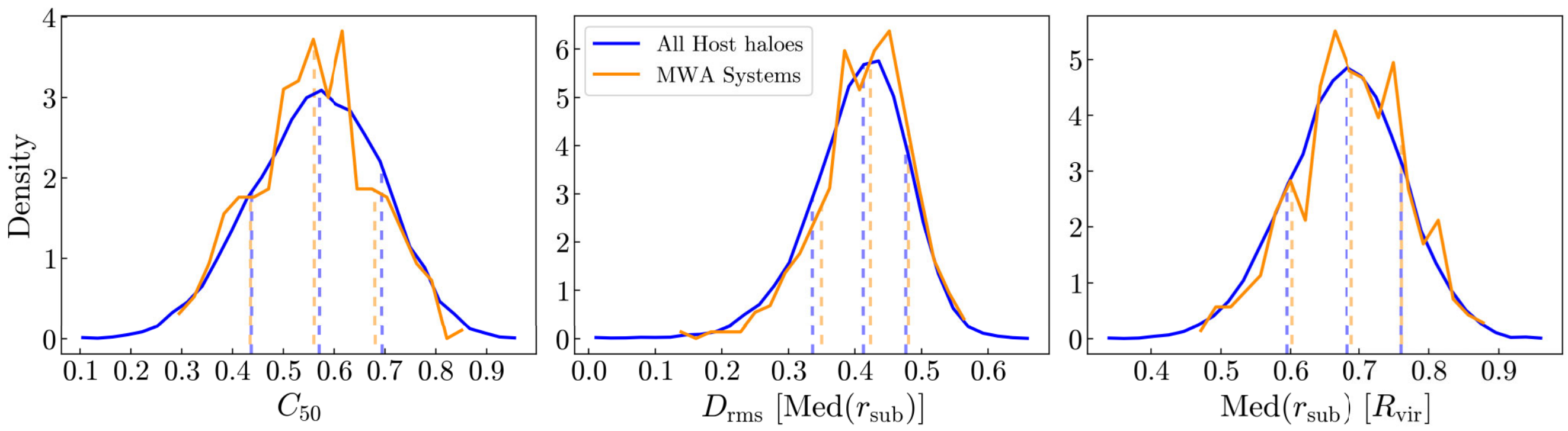

**Figure 10.** The distribution of the $C_{90}$, $D_{\rm rms}$, and Med($r_{\rm sub}$) marks for hosts in paired Milky Way-Andromeda (MWA) analogue systems (orange) compared to the distribution of all host haloes (blue). The dashed lines show the 16th, 50th, and 84th percentiles of their respective distributions. The two distributions are consistent with one another.

## 5 CONCLUSIONS

In this manuscript, we described a novel study of the degree to which host dark matter haloes (e.g. groups and clusters) cluster as a function of the anisotropy, alignment, and planarity of the subhalo populations which they carry. We began with a study of the anisotropy of the populations of subhaloes with respect to their host haloes. We focused on measurements of the alignment of subhaloes with the major axes of the mass distributions of their hosts and the planarity of subhaloes. This was motivated by a variety of considerations, including an exploration of the feasibility of studying such alignment- or anisotropy-dependent clustering in a statistically-large set of observational systems as a test of structure formation and galaxy evolution in the standard cosmological model.

We found that subhaloes are *not* distributed isotropically about their host haloes. Rather, subhaloes are distributed anisotropically about their hosts and tend to be aligned with the major axes of their host halo mass distributions. This confirms previous findings by a number of authors (e.g. see H. Y. Wang et al. 2005; A. R. Zentner et al. 2005b; N. I. Libeskind et al. 2005; N. I. Libeskind et al. 2007; N. I. Libeskind et al. 2011; M. S. Pawlowski et al. 2012; N. I. Libeskind et al. 2015; J. Shi et al. 2015; X. Kang & P. Wang 2015; S. Shao et al. 2018; P. Wang et al. 2019; L. Mezini et al. 2025; K. J. Kanehisa et al. 2025, and references therein for a number of examples), though we summarize and present these results in a novel way in Fig. 4. This figure also shows that subhaloes are distributed considerably more isotropically and exhibit less planarity than they would if they were tracers of their host halo triaxial ellipsoids. Subhaloes are more isotropic than dark matter.

We found for the first time that host halo systems cluster in a manner that depends upon the spatial distributions of their subhaloes. In particular, host systems in which the subhalo population is more *poorly aligned* with the mass distribution of the host halo *cluster more strongly*. Host haloes in which the subhaloes are distributed *less anisotropically* or exhibit the *least planarity* cluster *more strongly*. Moreover, we iterate on previous work and further show that host haloes in which their subhaloes reside preferentially at larger halocentric radii also cluster more strongly. In all cases, this dependence of clustering on subhalo distribution is statistically significant and large. We demonstrated these results using visual representations of the data, two-point correlation functions of halo subsamples, and, most powerfully, with a marked correlation function analysis.

We subsequently found no evidence that the dependence of halo clustering on subhalo alignment and radial position is *not* primarily induced by a secondary correlation between halo clustering and another individual halo property that also correlates with subhalo anisotropy. The clustering of host haloes as a function of the relative spatial distribution of their subhaloes, particularly alignment and radial position, may therefore be distinct from known secondary clustering dependencies and a novel effect.

This new result suggests a number of follow-up studies which we defer to future work. First, the large size of the anisotropy/alignment-dependent clustering effect suggests that it may be detectable in observational data as a way to study the way in which galaxies map onto haloes and subhaloes or even as a test of the standard model of cosmological structure formation. We will pursue this in a forthcoming paper. Secondly, the origin of this effect is not clear. It would be interesting to investigate the cause of the dependence of host halo clustering on the spatial distributions of their subhaloes.

## ACKNOWLEDGEMENTS

We thank Lorena Mezini and Atinç Çağan Şengül for useful discussions. We thank the anonymous referee for inspiring and informing the comparison to the ellipsoidal mock. This work was supported by the US National Science Foundation by Grant NSF PHY 2112723, by the Pittsburth Particle Physics,

Astrophysics, and Cosmology Center (Pitt PACC) at the University of Pittsburgh, and by the Dietrich School of Arts and Sciences at the University of Pittsburgh.

This research enlisted extensive use of Python and many packages, including JUPYTER (jupyter.org), MATPLOTLIB (J. D. Hunter 2007), NUMPY (C. R. Harris et al. 2020), SCIPY (P. Virtanen et al. 2020), and HALOTOOLS (A. P. Hearin et al. 2017).

The MultiDark Database used in this paper and the web application providing online access to it were constructed as part of the activities of the German Astrophysical Virtual Observatory as result of a collaboration between the Leibniz-Institute for Astrophysics Potsdam (AIP) and the Spanish MultiDark Consolider Project CSD2009-00064. The Bolshoi and MultiDark simulations were run on the NASA's Pleiades supercomputer at the NASA Ames Research Center. The MultiDark-Planck (MDPL) and the BigMD simulation suite have been performed in the Supermuc supercomputer at LRZ using time granted by PRACE.

## DATA AVAILABILITY

The raw halo catalogues used in this work are publicly available at cosmosim.org.

## APPENDIX A: CALCULATING THE ELLIPSOIDAL MOCK SAMPLES

In this appendix, we describe two methods for building ellipsoidal mock subhalo distributions. An obvious strategy is to apply an ellipsoidal stretch to the points in a spherically symmetric mock. Defining the two axis ratios $q = (b/a)_{\rm host}$ and $s = (c/a)_{\rm host}$, we can map the spherically symmetric coordinate of each subhalo $\boldsymbol{x}$ to a new coordinate

$$\tilde{\boldsymbol{x}} = (qs)^{-1/3}\begin{pmatrix}1 & 0 & 0\\ 0 & q & 0\\ 0 & 0 & s\end{pmatrix}\boldsymbol{x}. \tag{A1}$$

This method has the advantage that it produces a distribution that has axis ratios $q$ and $s$ (at least in the limit of a very large number of points) and its isodensity contours are those same ellipsoids. The Jacobian factor $(qs)^{-1/3}$ ensures that if the density of the spherical distribution on a surface of constant radial coordinate is $\rho(r)$, then the density of the ellipsoidal distribution on the ellipsoidal surface of major axis length $r$ is also $\rho(r)$. However, as it is a constant overall factor and we scale out absolute distances, it is not relevant to this work.

One feature of the preceding method is that it distorts the radial distribution of points. With (without) the Jacobian factor of $(qs)^{-1/3}$, it preferentially concentrates points at large (small) radii. This is a necessary consequence because it is not possible to produce an ellipsoidal point distribution that satisfies the requirements of ellipsoidal isodensity contours and a preserved radial distribution of points simultaneously. The fact that radial distributions are distorted makes the effect of scaling out distances different from our other methods for scaling out absolute distances (scaling by median or virial radius).

As an alternative, we present an algorithm for creating an ellipsoidal mock catalogue that has the desired axis ratios of $q$ and $s$ *and* preserves radial position. To achieve this, we first apply an anisotropic stretch transformation and then rescale radial distances so that the final distances are identical to the initial distances. The first step is to compute the intermediate coordinate $\bar{\boldsymbol{x}}$ by the stretch,

$$\bar{\boldsymbol{x}} = \begin{pmatrix}1 & 0 & 0\\ 0 & q^{\star} & 0\\ 0 & 0 & s^{\star}\end{pmatrix}\boldsymbol{x}. \tag{A2}$$

We then rescale lengths so that the final ellipsoidal coordinate which we use is

$$\tilde{\boldsymbol{x}} = |\boldsymbol{x}|\frac{\bar{\boldsymbol{x}}}{|\bar{\boldsymbol{x}}|}. \tag{A3}$$

The ellipsoidal distribution described by subhaloes with positions $\tilde{\boldsymbol{x}}$ defined in this way has precisely the same radial distribution of points as the original distribution. The disadvantage is that $q^{\star}$ and $s^{\star}$ are not the axis ratios which we seek to enforce, which are $q$ and $s$. Instead, $q^{\star}(q, s)$ and $s^{\star}(q, s)$ are functions of $q$ and $s$, which must be calibrated. Once calibrated, we can choose values of $q^{\star}$ and $s^{\star}$ that result in ellipsoids with axis ratios $q$ and $s$ after scaling the subhalo position vectors to their original length. A visual representation of this calibration is shown in Fig. A1.

The algorithm described in the preceding paragraph produces a mock catalogue of subhaloes distributed such that their principal axis ratios match $q$ and $s$ of the underlying dark matter distribution *and* preserves the radial distribution of subhaloes. What this algorithm does *not* achieve is that the isodensity contours are no longer ellipsoids with axis ratios $q$ and $s$. Along an ellipsoid of axis ratios $q$ and $s$, the density varies. This limitation in the construction of ellipsoidal models is known as is the fact that the dark matter distributions in dark matter haloes do not have exactly ellipsoidal isodensity contours with axis ratios that match axis ratios inferred from the shape tensor (e.g. Y. P. Jing & Y. Suto 2002; J. Bailin & M. Steinmetz 2005; E. Hayashi, J. F. Navarro & V. Springel 2007), but it is not known which of the algorithms described in this Appendix best represents the distributions of subhaloes for our purposes. For the sake of completeness, we present both of these methods in Fig. 4 where the first method is referred to as 'Ell. Mock $\rho$-iso' and shown in red while the second method is referred to as 'Ell. Mock $\Delta R_{\rm sub} = 0$' and shown in green.

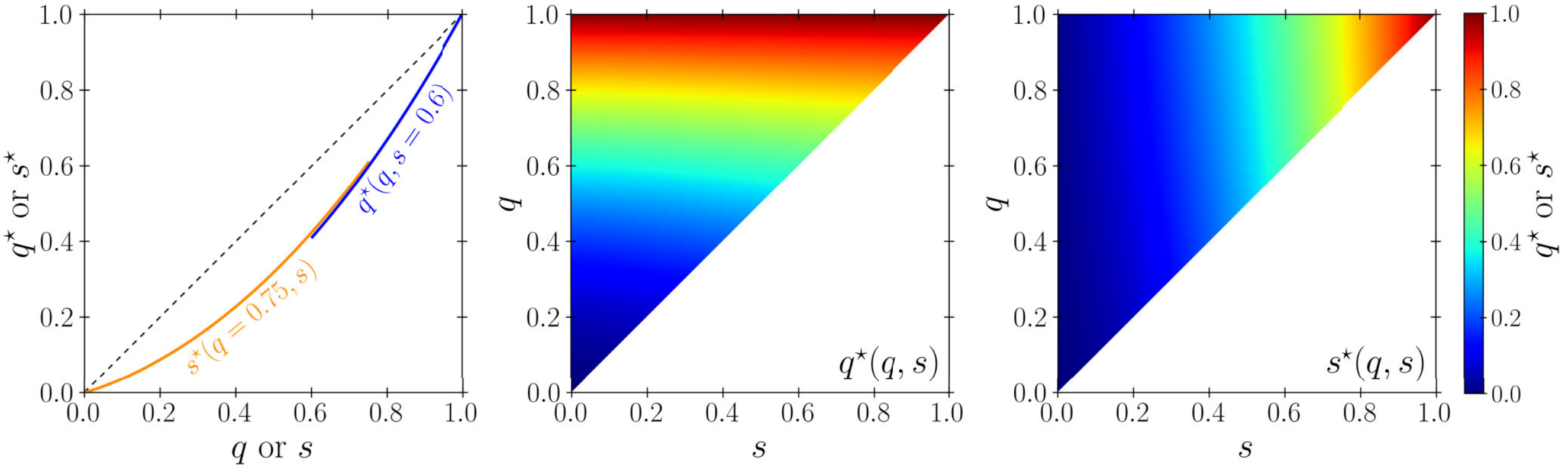

**Figure A1.** *Left:* An example of the calibration of the values of $q^\star$ and $s^\star$ needed to realize specific values of $q$ and $s$ as a function of $q$ and $s$. The lower, orange line shows $s^\star$ as a function of $s$ for a fixed value of $q = 0.75$ (so this line terminates at $s = 0.75$ because $s \le q$). The upper, blue line shows $q^\star$ as a function of $q$ for a fixed value of $s = 0.6$ and similarly terminates at $q = 0.6$. The black, dashed line is the line of equality which would mark $q^\star = q$ or $s^\star = s$. *Right:* Two panels showing $q^\star$ (left) and $s^\star$ (right) as a function of $q$ and $s$ for the entire range of possible values [0,1]. The color of the points in the panels signify the value of $q^\star$ and $s^\star$. These two panels showcase a family of functions of which the leftmost panel in this figure is an example. The lower right half of these panels are blank as this is the region where $s > q$.

Regardless of which method is used to produce an ellipsoidal mock distribution of subhaloes, we rotate the resulting mock ellipsoidal distribution of subhaloes about each host so that its principal axes (or the theoretical principal axes which would be obtained in the limit of an infinitely large sample) are aligned with the principal axes of the dark matter distribution of the host halo. In the mock systems which we realize, the ellipsoidal subhalo distributions are sampled by a finite number of subhaloes. Consequently, the measured subhalo principal axes are not guaranteed to align with the measured principal axes of the host dark matter distributions.

## APPENDIX B: THE DIRECT RELATIONSHIP BETWEEN SUBHALO ANISOTROPY AND KNOWN SECONDARY BIAS HALO PROPERTIES

In Section 4.1, we demonstrated how our measurement of subhalo anisotropy dependent clustering is independent of the known secondary biases of host halo concentration, spin parameter, host halo shape, and subhalo count. We justified those independent tests by saying that our marks are directly related to these secondary bias properties, here we discuss these relationships in detail.

Fig. B1 depicts subhalo alignment ($\widetilde{C}_{50}$), subhalo planarity ($\widetilde{D}_{\rm rms}$), and subhalo radial distribution ($\widetilde{{\rm Med}(r_{\rm sub})}$), as a function of host halo concentration ($\widetilde{c}_{\rm vir}$), spin ($\widetilde{\lambda}$), shape, ($\widetilde{(c/a)}_{\rm host}$, and subhalo count, $\widetilde{N}_{\rm sub}$ (all mass-normalized, as indicated by the tildes). The blue, solid lines in Fig. B1 are the median values in the bins, while the error bars are the standard error of the median for each bin. The light blue shaded region in each panel is the envelope spanning the 16$^{\rm th}$ and 84$^{\rm th}$ percentiles of each bin, making it a '1$\sigma$' envelope.

The top, left of panel in Fig. B1 shows a positive correlation between $\widetilde{C}_{50}$ and $\widetilde{c}_{\rm vir}$. The correlation is stronger at low concentration. This is quite interesting in the context of our results because low-concentration haloes cluster more strongly in our sample (Fig. 7). Recall that alignment as measured by $\widetilde{C}_{50}$ is such that less aligned (lower $\widetilde{C}_{50}$) hosts cluster more strongly as well. This suggests that the alignment-dependent clustering we reported in the previous section could, at least in part, be due to the correlation between alignment ($\widetilde{C}_{50}$) and concentration ($\widetilde{c}_{\rm vir}$).

The next panel in the top row of Fig. B1 shows a weak anticorrelation between $\widetilde{C}_{50}$ and spin, $\widetilde{\lambda}$ – higher spin host haloes have subhaloes with lower $\widetilde{C}_{50}$. As high-spin haloes cluster more strongly, just as low-alignment haloes do, this also suggests the possibility that the alignment-dependent clustering of the previous section could be induced by the combination of the correlation between spin and alignment with the known spin-dependent clustering of host dark matter haloes.

Proceeding across the top row of Fig. B1, the next panel displays a complicated relation between $\widetilde{C}_{50}$ and host halo shape. In this case, the correlation is opposite in sense to that which would be needed to explain the alignment-dependent clustering of our host systems through the correlation between host shape and alignment. More spherical (larger $\widetilde{(c/a)}_{\rm host}$) host haloes cluster more strongly, but they tend to host more well-aligned satellites (larger $\widetilde{C}_{50}$). Hosts with larger $\widetilde{C}_{50}$ tend to cluster more weakly.

Finally, the rightmost panel of the top row of Fig. B1 gives the alignment mark $\widetilde{C}_{50}$ as a function of subhalo count, $\widetilde{N}_{\rm sub}$. The correlation in this case is very weak. This indicates that the number of subhaloes does not strongly impact the measurement of $\widetilde{C}_{50}$, and that a correlation between these two variables does not explain alignment-dependent clustering.

The middle row of panels in Fig. B1 mimics the top row, but with our planarity measure $\widetilde{D}_{\rm rms}$, on the vertical axes of each panel. While most of the correlations are weak, particularly that between thickness and spin (second panel from left), they all have the sense that they could partly explain the subhalo planarity-dependent clustering of host systems. For example, consider the correlation between $\widetilde{D}_{\rm rms}$ and $\widetilde{N}_{\rm sub}$. Not only are these variables correlated, but this is the strongest correlation in the entire figure. Unsurprisingly, measurements of planarity are greatly influenced by the number of subhaloes that sample the spatial distribution.

Finally, the bottom row of Fig. B1 gives the correlation of the median subhalo position mark with (from left to right) concentration, spin, host halo shape, and subhalo count. Systems in which the subhaloes reside further from their hosts (large $\widetilde{\rm Med(}r_{\rm sub})$) have lower concentrations. Systems with higher spins have larger $\widetilde{\rm Med(}r_{\rm sub})$. Systems with that are more spherical have smaller $\widetilde{\rm Med(}r_{\rm sub})$. And systems with more subhaloes have slightly smaller median radial positions.

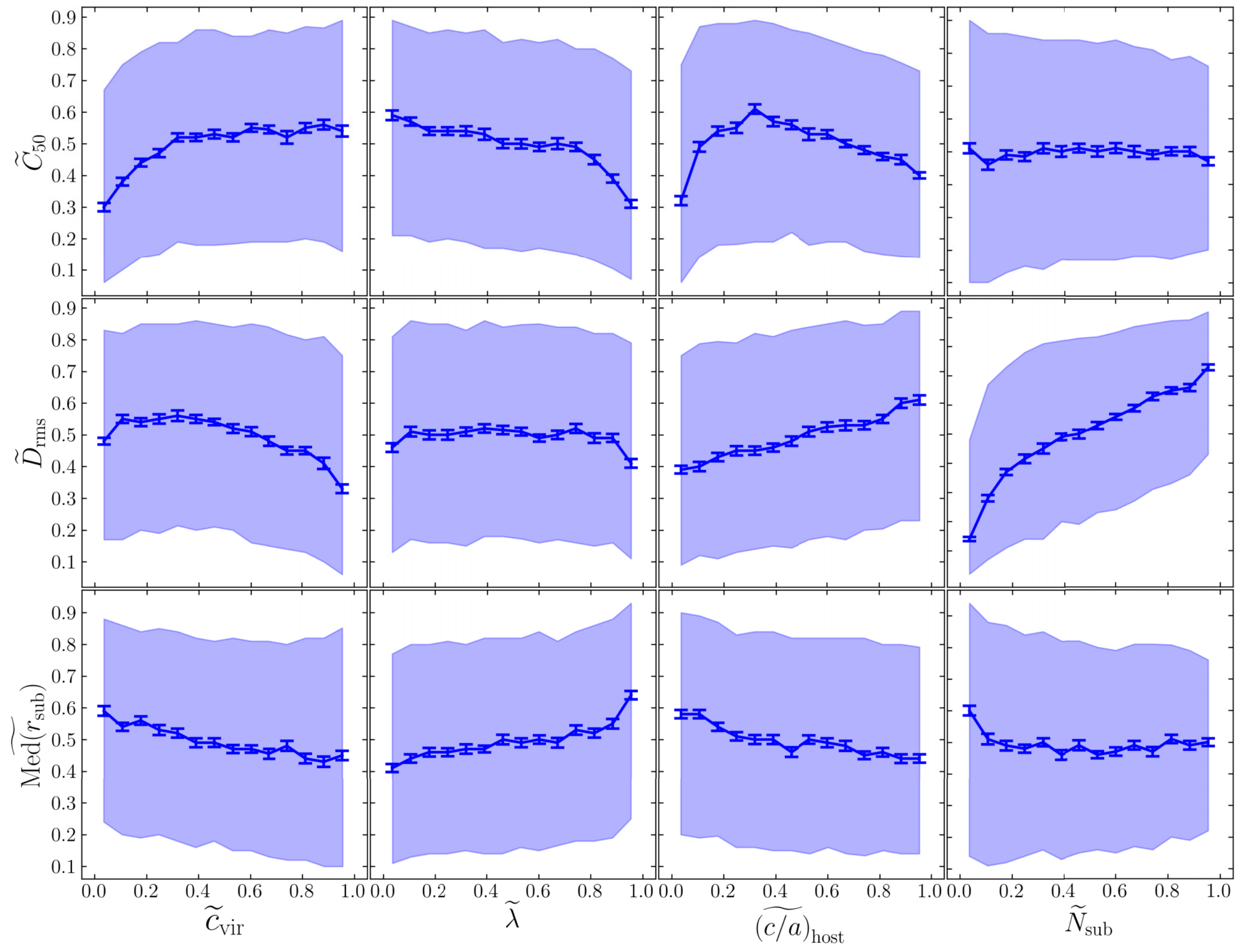

**Figure B1.** The relations between the subhalo spatial distribution marks $\widetilde{C}_{50}$, $\widetilde{D}_{\rm rms}$, and $\widetilde{\rm Med(}r_{\rm sub})$ and the halo properties concentration ($\widetilde{c}_{\rm vir}$), spin ($\widetilde{\lambda}$), host halo shape ($\widetilde{(c/a)}_{\rm host}$), and number of subhaloes ($\widetilde{N}_{\rm sub}$). These plots show the median of each bin as the thick line while the error bars are the standard error of the median. The shaded area is the envelope of each bin spanning the 16th to the 84th percentile, making it a '1$\sigma$' envelope. These panels show that the alignment, planarity, and radial distribution marks of host haloes are correlated with concentration, spin, host halo shape, and number of subhaloes.

## APPENDIX C: AN ADDITIONAL TEST OF THE RELATIONSHIP BETWEEN SUBHALO ANISOTROPY DEPENDENT CLUSTERING AND KNOWN SECONDARY BIASES

As an additional test of the relationship between subhalo anisotropy dependent clustering and the secondary biases of concentration, spin parameter, host halo shape, and subhalo count, we present Fig. C1. This figure is based off of Fig. 6 of Y.-Y. Mao et al. (2018), showing the direct relationship between our marks and the secondary bias properties, all mass normalized. In each panel, there are two subsamples of host's to compare. In *blue*, is the relationship for haloes that are not in a pair with another host and in *orange* is the relationship for host that are in a pairing. Following Y.-Y. Mao et al. (2018), our definition of paired versus unpaired is whether or not a host has a neighboring halo within 10 $h^{-1}$ Mpc. In Fig. C1, the lines show the median of each bin while the error bars show the standard error of the mean.

In each panel, a consistent difference between the two subsamples shows that the presence of a paired host halo is more impactful to the subhalo anisotropy mark than the value of secondary bias mark, and therefore independent of the secondary assembly bias. This effect can be seen clearly across the top and bottom rows which analyse the dependence for our $C_{50}$ and Med($r_{\rm sub}$) marks, indicating no evidence for a connection between subhalo alignment and radial distribution dependent clustering and these secondary biases. The middle row shows this exercise for our $D_{\rm rms}$ mark which exhibits a smaller difference between the two subsamples. This indicates a relationship between subhalo planarity dependent clustering and these secondary biases might be more present but a definitive conclusion cannot be made. This figure corroborates the conclusions drawn from Figs 8 and 9.

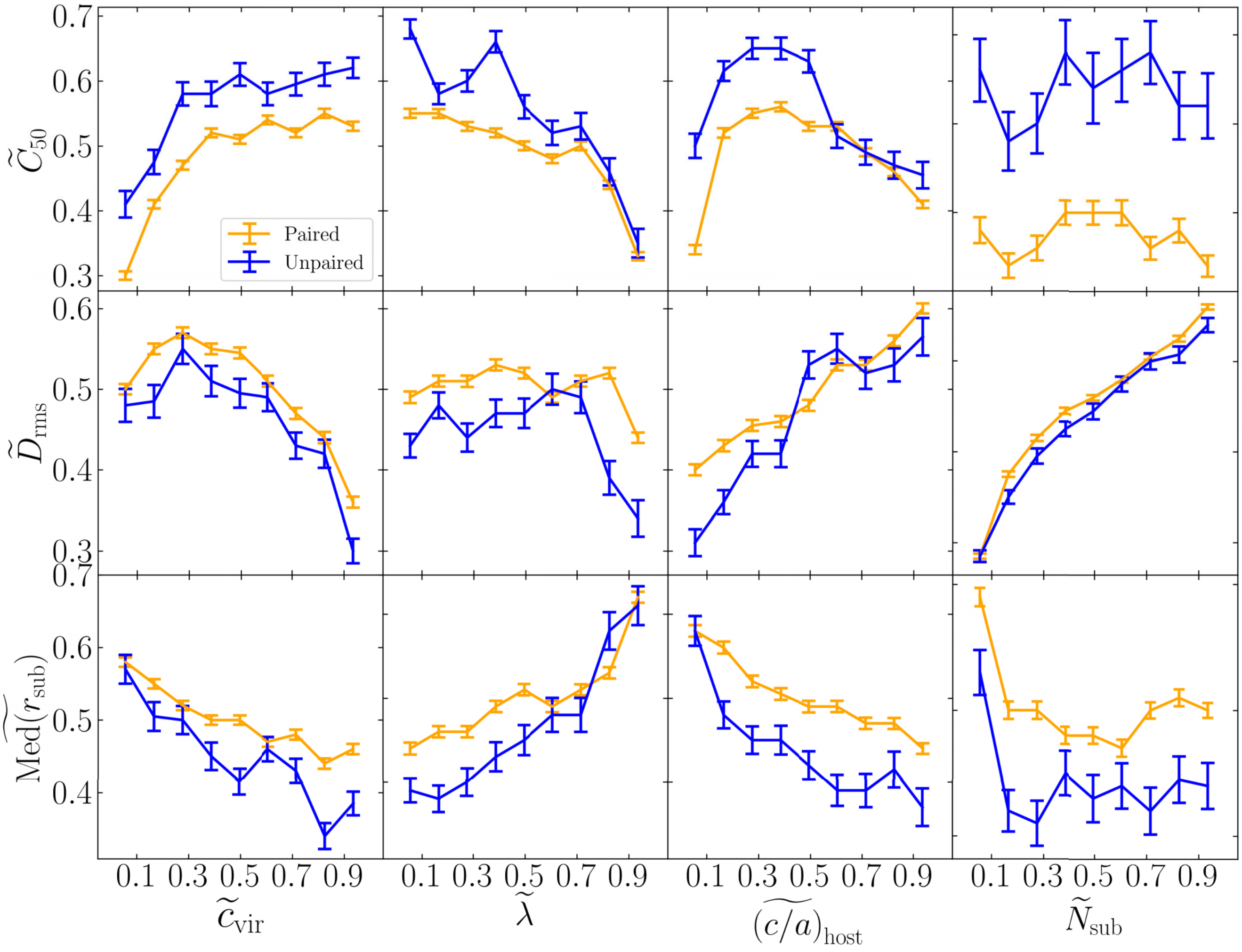

**Figure C1.** A comparison of the relationship between our marks and other known secondary biases when the host is in a paring (orange) versus when it is not (blue). This figure is based off of fig. 6 of Y.-Y. Mao et al. (2018) and illustrates the same conclusions of Figs 8 and 9.

This paper has been typeset from a TEX/LATEX file prepared by the author.